\documentclass[11pt]{article}
\usepackage{graphicx}
\usepackage{epsfig}
\usepackage{bbm}

\newcommand{\be}{\begin{eqnarray}}
\newcommand{\ee}{\end{eqnarray}}

\def\nue{{\nu_e}}

\def\numu{{\nu_{\mu}}}
\def\anumu{{\bar\nu_{\mu}}}
\def\nutau{{\nu_{\tau}}}

\newcommand{\ms}{\Delta m^2_{21}}
\newcommand{\ma}{\Delta m^2_{31}}
\newcommand{\meff}{\Delta m^2_{\rm eff}}

\newcommand{\sss}{\sin^2 \theta_{12}}

\newcommand{\stch}{\sin^2 2\theta_{13}}
\newcommand{\sa}{\sin^2 \theta_{23}}
\newcommand{\sta}{\sin^22 \theta_{23}}

\newcommand{\stcht}{\sin^2 2\theta_{13}{\mbox {(true)}}}
\newcommand{\sat}{\sin^2 \theta_{23}{\mbox {(true)}}}

\newcommand{\dcpt}{\delta_{CP}{\mbox {(true)}}}
\newcommand{\dcp}{\delta_{CP}}

\def\ltap{\ \raisebox{-.4ex}{\rlap{$\sim$}} \raisebox{.4ex}{$<$}\ }

\def\gs{\mathrel{
   \rlap{\raise 0.511ex \hbox{$>$}}{\lower 0.511ex \hbox{$\sim$}}}}
\def\ls{\mathrel{
   \rlap{\raise 0.511ex \hbox{$<$}}{\lower 0.511ex \hbox{$\sim$}}}}
\newcommand{\bea}{\begin{equation} \begin{array}{c}}
\newcommand{\bead}{\begin{equation} \begin{array}{cccc}}
\newcommand{\eea}{ \end{array} \end{equation}}

\begin{document}

\title{\bf Determining the Neutrino Mass Hierarchy with INO, T2K, NOvA and 
Reactor Experiments
}
\author{
Anushree Ghosh$^a$\,\thanks{email:\tt anushree@hri.res.in}~,~~
Tarak Thakore$^b$\,\thanks{email:\tt tarak@tifr.res.in}~,~~
Sandhya Choubey$^a$\,\thanks{email: \tt sandhya@hri.res.in}~,~~
\\\\
\\
{\normalsize \it$^{a}$Harish-Chandra Research Institute, Chhatnag Road, Jhunsi, Allahabad 211 019, India}\\ \\
{\normalsize \it$^{b}$Tata Institute of Fundamental Research, 1, Homi Bhabha Road, Mumbai 400 005, India}\\ \\
}
\date{\today}

\maketitle
\begin{abstract}\noindent

The relatively large measured value of $\theta_{13}$ has opened up the possibility of 
determining the neutrino mass hierarchy through earth matter effects. Amongst the 
current accelerator-based experiments only NOvA has a long enough baseline to 
observe earth matter effects. However, NOvA is plagued with uncertainty 
on the knowledge of the true value of $\delta_{CP}$, and this could drastically reduce its 
sensitivity to the neutrino mass hierarchy. The earth matter effect on 
atmospheric neutrinos on the other hand is almost independent of 
$\delta_{CP}$. The 50 kton magnetized Iron CALorimeter at the India-based 
Neutrino Observatory (ICAL@INO) will be observing atmospheric neutrinos. The charge 
identification capability of this detector gives it an edge over others 
for mass hierarchy determination through observation of earth matter effects. 
We study in detail the neutrino mass hierarchy sensitivity of the data from 
this experiment simulated using the NUANCE 
based generator developed for ICAL@INO and folded with the 
detector resolutions and efficiencies obtained by the INO collaboration 
from a full Geant4-based detector simulation. The data from ICAL@INO 
is then combined with simulated data from T2K, NOvA, 
Double Chooz, RENO and Daya Bay 
experiments and a combined sensitivity study to the mass hierarchy is 
performed. With 10 years of ICAL@INO data combined with T2K, NOvA
and reactor data, one could get about $2.3\sigma-5.7\sigma$ discovery of the neutrino 
mass hierarchy, depending on the true value of $\sin^2\theta_{23}$ [0.4 -- 0.6], 
$\sin^22\theta_{13}$ [0.08 -- 0.12] and 
$\delta_{CP}$ [0 -- 2$\pi$].

\end{abstract}

\newpage

\section{Introduction}

The past year 
has been very exciting in the field of neutrino physics. Results from as many as five 
different experiments finally confirmed that the mixing angle $\theta_{13}$ is 
non-zero and indeed {\it large}\footnote{``Large" here implies that the measured value of 
$\theta_{13}$ turned out to be larger than expected. In fact, it was found to lie just below 
the earlier upper bound on this parameter. However, it is still smaller compared to the 
other two mixing angles $\theta_{23}$ and $\theta_{12}$.}. 
The first hints came in 2011 from the T2K experiment \cite{t2kth13},
which with just 6 events collected before its forced temporary halt due to the great Japan earthquake, 
inferred a value of $\stch \simeq 0.1$, with a zero value excluded at the $2.5\sigma$ C.L.. This was 
corroborated with similar hints from the MINOS \cite{minosth13} and the Double Chooz \cite{dcth13} 
experiments. Finally, the Daya Bay experiment put all doubts to rest in March 2012, when with 
just 55 live days of their data they 
established a non-zero $\theta_{13}$ at the $5.2\sigma$ C.L. with the best-fit 
$\stch = 0.092 \pm 0.016({\rm stat}) \pm 0.005 ({\rm syst})$ 
\cite{dbth13}. These results were further strengthened by the RENO 
experiment which provided an independent confirmation of $\theta_{13} \neq 0$ at the 
$4.9\sigma$ C.L. in April 2012, with a best-fit 
$\stch  = 0.113 \pm 0.013({\rm stat}) \pm 0.019 ({\rm syst})$ \cite{renoth13}. All these experiments 
confirmed their first results with more statistics at the Neutrino Conference in Kyoto, Japan \cite{nu2012}. 
A global analysis of the world neutrino data now provides a more than $10\sigma$ signal for a non-zero 
$\theta_{13}$, with best-fit $\stch \simeq 0.1 \pm 0.01$ 
\cite{thomas_ggi,globallisi,globalvalle}. 

The relatively {\it large} value of $\theta_{13}$ has opened up the possibility of 
answering the   
two other major elusive issues in neutrino 
oscillation physics -- CP violation in the lepton sector, and 
the sign of $\ma$, {\it aka}, the neutrino mass hierarchy\footnote{We define mass squared 
differences as $\Delta m^2_{ji} = m_j^2 - m_i^2$.}.  
In order to address these two issues 
a huge effort worldwide is underway and next generation experiments are being 
proposed. 
In the light of the large measured value of $\stch$, one 
needs to take another look at the physics reach of these prospective experimental endeavors 
to allow one to make prudent choices for the design of the next generation long baseline 
experiments. While the large value of $\stch$ would necessarily improve the prospects of 
determining the neutrino mass hierarchy through observations of earth matter effects in neutrino 
beam experiments, 
the same is not necessarily true for the measurement of CP violation \cite{minakata2012}. 
The latter effect 
appears through the sub-dominant term in the $\numu \to \nue$ oscillation probability, and hence is  
much more subtle. 
Therefore, it is pertinent to ask at this point if one could measure the 
neutrino mass hierarchy elsewhere \cite{efm2012}. 
One could then optimize the long baseline experiment 
mainly to measure CP violation. In particular, 
with $\stch \simeq 0.1$ one could use the earth matter effects 
in atmospheric neutrinos in order to look for, and possibly identify the 
neutrino mass hierarchy. 

Prospects of detecting earth matter effects in atmospheric neutrinos have been 
discussed in detail in the past 
by many authors \cite{petcov2002}-\cite{bs2012}. 
In particular, the possibility of measuring the neutrino mass hierarchy in 
magnetized iron calorimeters has been considered before. In this paper, we 
expound the reach of atmospheric neutrino measurements at 
the magnetized Iron CALorimeter (ICAL) detector to be built at the 
India-based Neutrino Observatory (INO) for the determination of the mass hierarchy. 
Henceforth we will designate this experiment as ICAL@INO. 
The main new elements in this work are the simulation of atmospheric neutrino events in 
ICAL@INO and its detailed physics analysis,  
for which we use the tools developed 
by the INO collaboration
for this experiment. Most past works on the mass hierarchy determination with 
magnetized iron calorimeters performed the analysis in terms of the neutrino energy and 
zenith angle bins, with some assumed fixed values for the 
resolutions and efficiencies. In this work, we perform our analysis in terms of the muons,  
which are binned in 
reconstructed energy and zenith angle bins. 
For obtaining the reconstructed energy and zenith angle of the 
muons we use, for the first time, the detailed 
efficiencies and resolution functions obtained by the INO collaboration from 
the simulation codes developed for ICAL. 
In particular, 
we use the NUANCE V3.000 \cite{nuance} based atmospheric neutrino event generator used by the 
INO collaboration to simulate the events in the detector in the absence of 
oscillations. This raw NUANCE data is 
then folded with the oscillation code via a  reweighting algorithm 
and the oscillated event spectrum generated. 
These events are then 
folded with the resolutions and efficiencies obtained by 
the INO collaboration from a full Geant4-based detector Monte Carlo 
developed to simulate the ICAL@INO detector \cite{inomuon}. Separate papers will 
appear describing in detail the simulation codes and the results from these
studies performed by the collaboration \cite{inomuon,inohadron}. Here we use these results to 
do the physics analysis of simulated atmospheric neutrino events in 
ICAL@INO. In particular, we study the reach 
of the experiment in pinning down the true neutrino mass hierarchy as 
a function of the number of years of running of the experiment. We quantitatively 
show how this sensitivity depends on the uncertainties in the measurement of the 
other oscillation parameters, especially $|\Delta m^2_{31}|$, $\theta_{23}$ and 
$\theta_{13}$. Since all three of these are expected to be measured to a remarkable 
precision by the current reactor and accelerator-based long baseline experiments, 
we include in our analysis the simulated data from the full run of all of these 
experiments and show the joint sensitivity reach to the neutrino mass hierarchy. 
We find that while these reactor and accelerator-based neutrino 
oscillation experiments themselves have 
very limited sensitivity to the neutrino mass hierarchy, they still have a crucial role 
to play in this effort, since they constrain $|\Delta m^2_{31}|$, $\theta_{23}$ and 
$\theta_{13}$, which in turn improves the statistical significance with which 
ICAL@INO can determine the neutrino mass hierarchy. We also show that 
the reach of ICAL@INO for the neutrino mass hierarchy is nearly independent 
of $\delta_{CP}$, which at the moment is totally unknown and which is also the 
hardest amongst all oscillation parameters to be measured. The long baseline 
experiments, on the other hand are bogged down by the uncertainty 
in the true value of $\delta_{CP}$, making it difficult to measure the neutrino 
mass hierarchy from these experiments alone \cite{tenyrs09}. We will show that 
ICAL@INO will provide a remarkable 
complementarity in this direction.

The paper is organized as follows. In section 2 we briefly review the current status 
of the already measured neutrino oscillation parameters and give their 
benchmark values at which we simulate the projected data used in our analysis. 
We give the ranges of these parameters over which they are allowed to 
vary in our statistical fits. In section 3 we describe the experiments 
ICAL@INO, Double Chooz, Daya Bay, RENO, T2K, and NO$\nu$A and 
spell out the experimental specifications used in our analysis for each one of them. 
In section 4 we present the simulated events at ICAL@INO and show the 
impact of the efficiencies and resolutions for the muons in ICAL obtained by 
the INO collaboration. In section 5 we give the details of our statistical analysis 
tool. In section 6 we present our main results. The impact of systematic uncertainties on 
the mass hierarchy sensitivity of ICAL@INO is discussed in section 7 and that of  
the true value of $\theta_{23}$ is studied in section 8. In section 9 we explore the 
effect of $\delta_{CP}$ value on the mass hierarchy sensitivity in ICAL@INO and 
NO$\nu$A. 
We end with our conclusions in 
section 10.

\section{Neutrino Oscillation Paramaters}

\begin{table}[h]
\begin{center}
\begin{tabular}{l|c|c}
\hline
&\\[-0.9mm]
Parameter & True value used in data & $3\sigma$ range used in fit \\[2mm] \hline
&&\\
[-0.9mm]$\ms$ & $7.5 \times 10^{-5}$ eV$^2$ & $[7.0-8.0] \times 10^{-5}$ eV$^2$\\[2mm]
$\sin^2\theta_{12}$ & 0.3 & $[0.265-0.33]$\\[2mm]
$|\meff|$ & $2.4 \times 10^{-3}$ eV$^2$ & $[2.1-2.6]\times 10^{-3}$ eV$^2$\\[2mm]
$\delta_{CP}$ & 0 & $[0 - 2\pi]$\\[2mm]
\hline
&&\\
[-0.9mm]$\sa$ & 0.4, 0.5, 0.6 & $\sin^2\theta_{23}$(true)$\pm 0.1$  \\[2mm]
$\stch$ & 0.08, 0.1, 0.12 & $\stch$(true)$\pm 0.03$   \\[2mm]
\hline
\end{tabular}
\caption{\label{tab:param}
Benchmark true values of oscillation parameters 
used in the simulations, unless otherwise stated. The range over which they are 
allowed to vary freely in the fit is also shown in the last column. For $\sat$ and 
$\stcht$ we use three benchmark values for simulating the data. 
}
\end{center}
\end{table}

We start with a brief overview of the current status of the neutrino oscillation 
parameters \cite{thomas_ggi,globallisi,globalvalle}
and their best-fit values which we use for simulating events 
in the various experiments. The solar neutrino parameters 
$\ms$ and $\sin^2\theta_{12}$ are now 
determined to extremely good precision by the joint analysis of the solar \cite{solar} and 
KamLAND \cite{kl} data. Observation of matter effects in solar neutrinos has also 
nailed down the sign of $\ms$ to be positive. 
The current best-fit from a 
global analysis \cite{globallisi} is 
\be
\ms = (7.54 \pm 0.26) \times 10^{-5} eV^2,\,\,\, \sin^2\theta_{12} =0.31\pm 0.02
\,.
\ee
Among the atmospheric neutrino 
parameters, $\sta$ is mainly constrained by the Super-Kamiokande atmospheric 
neutrino data \cite{sk}, while $|\ma|$ is predominantly determined by the MINOS long baseline 
disappearance data \cite{minos}. For $|\ma|$ the current 
best-fit is close to \cite{thomas_ggi,globallisi}
\be
|\ma| = 2.47 \times 10^{-3} eV^2
\,.
\ee
With neutrino physics entering the precision era, it has become very important to define 
what is meant by the atmospheric neutrino mass splitting when one is doing a three-generation 
fit. 
The subtlety involved is the following. The value of the best-fit for the 
atmospheric mass squared difference depends on the mass hierarchy and definition used.
In particular, it could be rather misleading to use an inconsistent definition for this 
parameter when doing mass hierarchy studies. For instance, if $\ma$ is called the 
atmospheric neutrino mass squared difference and $\ma >0$ defined as normal hierarchy, 
then the absolute value of 
$\Delta m^2_{32}$ changes when one changes the hierarchy from normal ($\ma>0$) to 
inverted ($\ma <0$). Since the three generation oscillation probability is sensitive to all 
oscillation frequencies, in this case one gets a rather large difference in the 
survival probability $P_{\numu\numu}$ between normal and inverted hierarchies 
even when $\theta_{13}$ is taken as zero and there are no $\theta_{13}$ driven earth matter effects. 
One needs to perform a careful marginalization over $\ma$ in this case to get rid of the 
spurious difference in $P_{\numu\numu}$ coming from this effect \cite{gandhizero}. 
Therefore, it is important to use a consistent definition for the mass squared 
difference in the analysis, especially in studies pertaining to observations of earth matter effects. 
In our study we use as the atmospheric mass squared difference, the quantity defined as \cite{meff}
\be
\meff = \ma - (\cos^2\theta_{12} - \cos\delta_{CP}\sin\theta_{13}\sin2\theta_{12}\tan\theta_{23})\ms
\,,
\label{eq:meff}
\ee
where the other parameters are defined according to the convention used by the PDG. 
The normal hierarchy is then defined as $\meff >0$ and the inverted hierarchy as $\meff <0$. 
Defining the mass squared difference by Eq. (\ref{eq:meff}) is 
particularly convenient for mass 
hierarchy studies involving the 
probability $P_{\numu\numu}$ since it is almost same for 
$\meff >0$ (normal hierarchy) and $\meff <0$ (inverted hierarchy) for $\theta_{13}=0$. 
This ensures that there is no spurious contribution to the mass hierarchy sensitivity coming 
from the difference between the oscillation frequencies for the normal and inverted 
hierarchies in experiments predominantly sensitive to the oscillation channel $P_{\numu\numu}$. 
However for this definition of the mass hierarchy, 
there is a difference between the 
frequencies involved in normal and inverted ordering 
for the other oscillation channels and marginalizing over 
$\meff$ then becomes very important for them. In our analysis we have paid 
special attention to the marginalizing procedure and have checked our mass 
hierarchy sensitivity to the definition used for the atmospheric mass squared difference and 
the mass hierarchy. We will quantify this issue later.

The issue regarding the value of $\sa$ and its correct octant is 
not yet settled. The Super-Kamiokande collaboration get the best-fit
to their atmospheric zenith angle data at  
$\sta=0.98$ \cite{sk}. The MINOS collaboration best-fit 
$\sta = 0.96$ \cite{minos}, where they included in their analysis the full MINOS data with 
$10.71\times 10^{20}$ POT for $\numu$-beam, $3.36\times 10^{20}$ POT for  $\anumu$-beams, 
as well as their 37.9 kton-years data from atmospheric 
neutrinos. It is worth pointing out here that the Super-Kamiokande best-fit is close to maximal and 
even the MINOS collaboration results allow maximal mixing at $1\sigma$ C.L.  \cite{minos}.
However, results from global analyses performed by groups outside 
the experimental collaborations have now 
started to show deviation from maximal mixing and prefer $\sa$ in the first octant 
\cite{globallisi,thomas_ggi}. 
These results though 
should be taken only as a possible ``hint" as they require further investigation by 
the analysis using the full detector Monte Carlo of the experimental collaborations. 

Our analysis in this paper uses the full three generation oscillation probabilities without 
any approximations. The data are simulated at a particular set of benchmark values chosen for the 
oscillation parameters, which we call ``true value". These are summarized in Table \ref{tab:param}. 
The true values of $\ms$, $\sss$ and $|\meff|$ are kept fixed throughout 
the paper. These quantities are now fairly well determined by the current global 
neutrino data and we choose our benchmark true values for these parameters to be  
close to their current best-fits, as discussed above. 
We will show results for a range of 
plausible values of $\sat$ and $\stcht$ since the 
earth matter effects are fairly sensitive to these parameters. 
While we have absolutely no knowledge on the value of $\delta_{CP}$(true), 
the sensitivity of ICAL@INO 
does not depend much on the true value of this parameter. 
Therefore, $\delta_{CP}$(true) is also kept fixed at zero, unless otherwise stated. 
Only in section 8, where we study the impact of $\delta_{CP}$(true) on the 
mass hierarchy reach of  
the NO$\nu$A experiment, will we show results as a function of the $\delta_{CP}$(true). 

In our fit, we allow the oscillation parameters to vary freely within their current 
$3\sigma$ limits and the $\chi^2$ is minimized (marginalized) over them. 
The range 
over which these parameters are varied in the fit is shown in Table \ref{tab:param}. 
Since the 
ICAL@INO sensitivity does not depend much on $\ms$, $\sss$ and $\delta_{CP}$, 
we keep these fixed at their true values in the fit for the analysis of the 
ICAL@INO data. However, the ICAL@INO sensitivity to the mass hierarchy 
does depend on 
$|\meff|$, $\sa$ and $\stch$ and 
hence the $\chi^2_{ino}$ is marginalized over the $3\sigma$ ranges 
of these parameters. 
The combined $\chi^2$ for the accelerator and reactor experiments depends 
on all the oscillation parameters, and so for them we marginalize the 
$\chi^2$ over the current $3\sigma$ range of all the oscillation parameters.
Since the range of $|\meff|$, $\sa$ and $\stch$ will be severely constrained 
by the future accelerator and reactor data themselves, the best-fit for these in 
our global fits are mostly close to the true value taken for the data. However, 
none of the data sets included in our analysis has the potential to 
constrain $\ms$ and $\sss$. Therefore, in order to take into account the fact 
that not all values of $\ms$ and $\sss$ within their current $3\sigma$ range 
are allowed with equal probability by the solar and KamLAND data, we impose a 
``prior" according to the following definition:
\be
\chi^2_{prior} = \sum_i\bigg(\frac{p_i^{fit} - p_i^{true}}{\sigma_{p_i}}\bigg)^2
\,,
\label{eq:prior}
\ee
where the parameter $p_i$ is $\ms$ and $\sss$ for $i=1$ and 2 respectively, with 
$\sigma_{\ms} = 3\%$ and $\sigma_{\sss}=4\%$. This $\chi^2_{prior}$ is added to 
the sum of the $\chi^2$ obtained from the analysis of the ICAL@INO simulated data and 
the prospective accelerator and reactor 
data, and the combined $\chi^2$ is then marginalized over all 
oscillation parameters.

\section{Neutrino Oscillation Experiments}

We give below a very brief description of the experiments whose simulated data 
we use in this analysis to pin down the neutrino mass hierarchy. The ICAL@INO 
experiment is of course the focus of this work. We begin with a short overview of 
this experimental endeavor. We then move on to mention just the key features of 
the accelerator-based experiments T2K and NO$\nu$A and the reactor experiments 
Double Chooz, Daya Bay and RENO. 

\subsection{ICAL@ India-based Neutrino Observatory}

ICAL will be a 50 kton magnetized iron calorimeter at the INO laboratory in India  
and will soon go into construction in the Theni district in Southern India.
It will be solid rectangular in shape with 
dimensions 48.4 m in length, 16 m in width and 14.4 m in height and will consist of   
three identical 
modules. The
detector will have a layered structure with 150 layers of 
5.6 cm iron slabs interleaved with glass 
Resistive Plate Chambers (RPC) acting as the active detector element. 
Each glass RPC to be deployed in ICAL@INO 
will be about 2 m $\times$ 2 m in size made up of two parallel glass electrodes separated 
by spacers to create a gap which is filled with a gas mixture of tetraflouroethane  
and isobutane. 
This particular combination of gases chosen by the INO collaboration 
enables the RPC to be used in the avalanche mode wherein 
the arrival of a charged particle results in a Townsend avalanche 
through the gas volume. 
A total of 
about 28,000 units of such RPCs will be required for the complete detector. 
Eight such units will be arranged next to each other to form 16 m $\times$ 2 m road and 
each module will have eight such roads per layer.  
This signal in the RPC will be read by 3 cm wide 
pickup strips that will be laid orthogonal to each other (X and Y strips). The signal will 
then go to front end ASICs located at the end of the strips. There will be 
64 strips long x direction and 64 strips along the y
direction per RPC. 
Thus one needs 
about 3.7 million electronic readout channels for the full detector. 
The orthogonal X-Y readout strips form a grid such that 
muons traveling in the detector will trigger the RPCs in a particular 3 cm $\times$ 3 cm block which 
serves as a hit point. 
Since the muon will cross a number of RPCs, the hit points can be joined 
to reconstruct the long well defined muon track. In addition, being made of iron, 
the ICAL detector will be magnetized with a magnetic field strength of about 1.3 Tesla.  
Therefore as it travels, the $\mu^-$ bends in a direction opposite to that of the $\mu^+$. 
This gives ICAL an edge over other detectors since the magnetic field allows charge 
discrimination allowing it to distinguish between 
muon neutrinos and antineutrinos. 
On the other hand, 
ICAL is expected to not have very good
sensitivity to electrons
since the electron showers 
are mostly absorbed in the dense iron material. However, hadron showers can be seen 
in the detector and their energy measured. This makes this detector a calorimeter 
wherein the energy and momentum of the incoming (anti)neutrino can be reconstructed 
by adding the energy and momentum of the resultant muon and hadron(s). 
Therefore the energy and momentum resolution of ICAL is expected to be better 
than that of detectors which are insensitive to the energy and momentum of the 
final state hadrons. 
At least four types of large detectors for atmospheric neutrinos have been 
envisaged. While magnetized iron calorimeters such as ICAL@INO have excellent 
charge identification capabilities and good energy and angle resolution, they suffer from difficulty in 
observing electrons and have a higher energy threshold for muons. The water Cherenkov detectors 
like Hyper-Kamiokande \cite{hk}
observe both muons and electrons with very low energy threshold, but 
cannot be magnetized. Liquid argon detectors \cite{gandhi2012,lar} 
have extremely good 
detector response but magnetization could still be a challenge for them \cite{larmag}. 
The multi-megaton ice detector 
PINGU (IceCube extension for low-energy neutrinos) \cite{pingu}
could use its huge statistics to overcome its other drawbacks to return 
good sensitivity to the mass hierarchy \cite{pinguatm}. 
The simulation of events in ICAL will be discussed in detail in the 
next section.


\subsection{Current Reactor and Accelerator Experiments}
\label{sec:lbl}

For simulation of the current reactor and accelerator-based experiments we use the 
GLoBES software \cite{globes}. We have closely followed \cite{tenyrs09} for 
the analysis of the future accelerator and 
reactor data. The experiments that we include in our study are the 
following:

\begin{itemize}
\item {\bf Double Chooz:} The Double Chooz reactor experiment \cite{dc} has a 
liquid scintillator detector with fiducial mass of 8.3 tons placed at a distance of 
1 km and 1.1 km from the two reactor cores of the Chooz reactor power plant, 
each with 4.27 GW$_{th}$ thermal power. Double Chooz has been taking data 
with just this far detector and 
have observed a positive signal for $\theta_{13}$ at $3.1\sigma$ C.L. \cite{dcth13,nu2012}. 
In addition to the far detector, this experiment will also have a near detector which 
will be identical to the far detector and will be placed at a distance of 
470 m and 350 m respectively from the two reactor cores. 
Following 
\cite{tenyrs09}, we perform the analysis 
for an exposure of 3 years with both near and far detectors fully operational and 
with detector efficiency of 80\% and reactor load factor of 78\%. An uncorrelated 
systematic uncertainty of 0.6\% is assumed.

\item {\bf RENO:} The RENO antineutrino experiment \cite{reno} is powered by the 
Yonggwang reactor plant in South Korea, with a total reactor power of 
16.4 GW$_{th}$, making it currently the most powerful reactor in the world behind 
the Kashiwazaki-Kariwa power plant in Japan (which is currently shutdown). 
This reactor complex has 
six reactor cores with 
first two having power 2.6 GW$_{th}$ while the last four with power 2.8 GW$_{th}$, 
respectively. These reactor cores are 
arranged along a 1.5 km straight line separated from each other by equal distances. 
The near detector of this experiment has a fiducial mass of 15 tons and is situated 
at a distance of 669 m, 453 m, 307 m, 338 m, 515 m and 74 m respectively from the reactor cores. 
The far detector also has a fiducial mass of 15 tons and 
is placed in the opposite direction 
at a distance of  1.557 km, 1.457 km, 1.396 km, 1.382 km, 1.414 km and 
1.491 km respectively from the reactor cores. 
The first data set from this experiment was released in March 2012 \cite{renoth13} 
confirming that $\theta_{13}$ was indeed non-zero at $4.9\sigma$ C.L.. 
We consider in our analysis simulated data with 3 years of full run for the 
RENO experiment and include an uncorrelated systematic uncertainty of 
0.5\% which is the projected benchmark systematic uncertainty for RENO \cite{nu2012}.  

\item {\bf Daya Bay:} The Daya Bay reactor experiment \cite{db}  
observes antineutrinos from the Daya Bay and Ling Ao I and Ling Ao II reactors. 
Each of them have two reactor cores with a total combined power of 17.4 GW$_{th}$. 
This experiment when fully constructed will have 4 near detectors each with 
20 ton fiducial mass and 4 far detectors also with fiducial mass 20 ton each. 
The far detectors will be at a distance of 1.985, 1.613 and 1.618 km respectively from 
Daya Bay, Ling Ao and Ling Ao II reactors. The distance of the near detectors for each of the 
reactor cores is more complicated, and can be found in \cite{db}. 
The experiment has been running with 6 detectors and 
so far produced outstanding results \cite{dbth13,nu2012}. 
We analyse simulated data corresponding to 3 years of full run for the 
Daya Bay experiment and consider an uncorrelated systematic uncertainty of 
0.18\% which is the projected systematic uncertainty for Daya Bay. 

\item {\bf T2K:} In the T2K experiment \cite{t2k} 
a 2.5$^\circ$ off-axis neutrino beam is sent from the 
J-PARC accelerator facility at Tokai to the 
Super-Kamiokande detector at Kamioka at a distance 
of 295 km. The beam power used for the simulation 
is taken to be 0.75 MW with 5 years of neutrino running. 
The fiducial mass of Super-Kamiokande is 22.5 kton and there is a near 
detector ND280 at a distance of 280 m from the beam target. 

\item{\bf NO$\nu$A:} The NO$\nu$A experiment \cite{nova}
will shoot a  3.3$^\circ$ off-axis (anti)neutrino beam from 
NuMI at Fermilab to the 15 kton Totally Active Scintillator Detector (TASD) 
located in Northern Minnesota at a distance of 810 km. 
The near detector at Fermilab is a 200 ton detector similar to 
the far detector. The beam power used is 0.7 MW with 
3 years of running in the neutrino and 3 in the antineutrino mode.

\end{itemize}


\section{Simulated Events in ICAL@INO}

\begin{figure}
\centering
\includegraphics[width=0.49\textwidth]{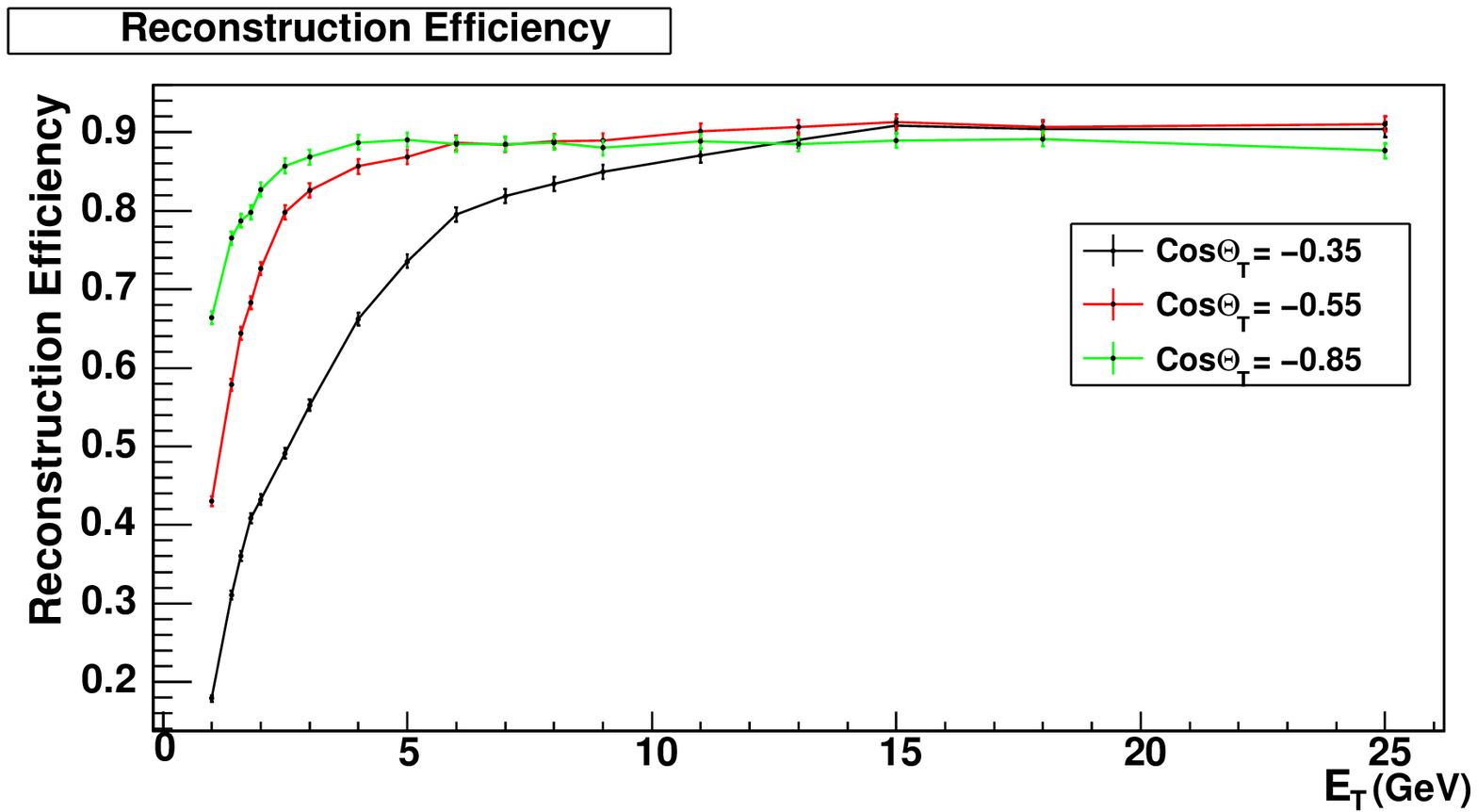}
\includegraphics[width=0.49\textwidth]{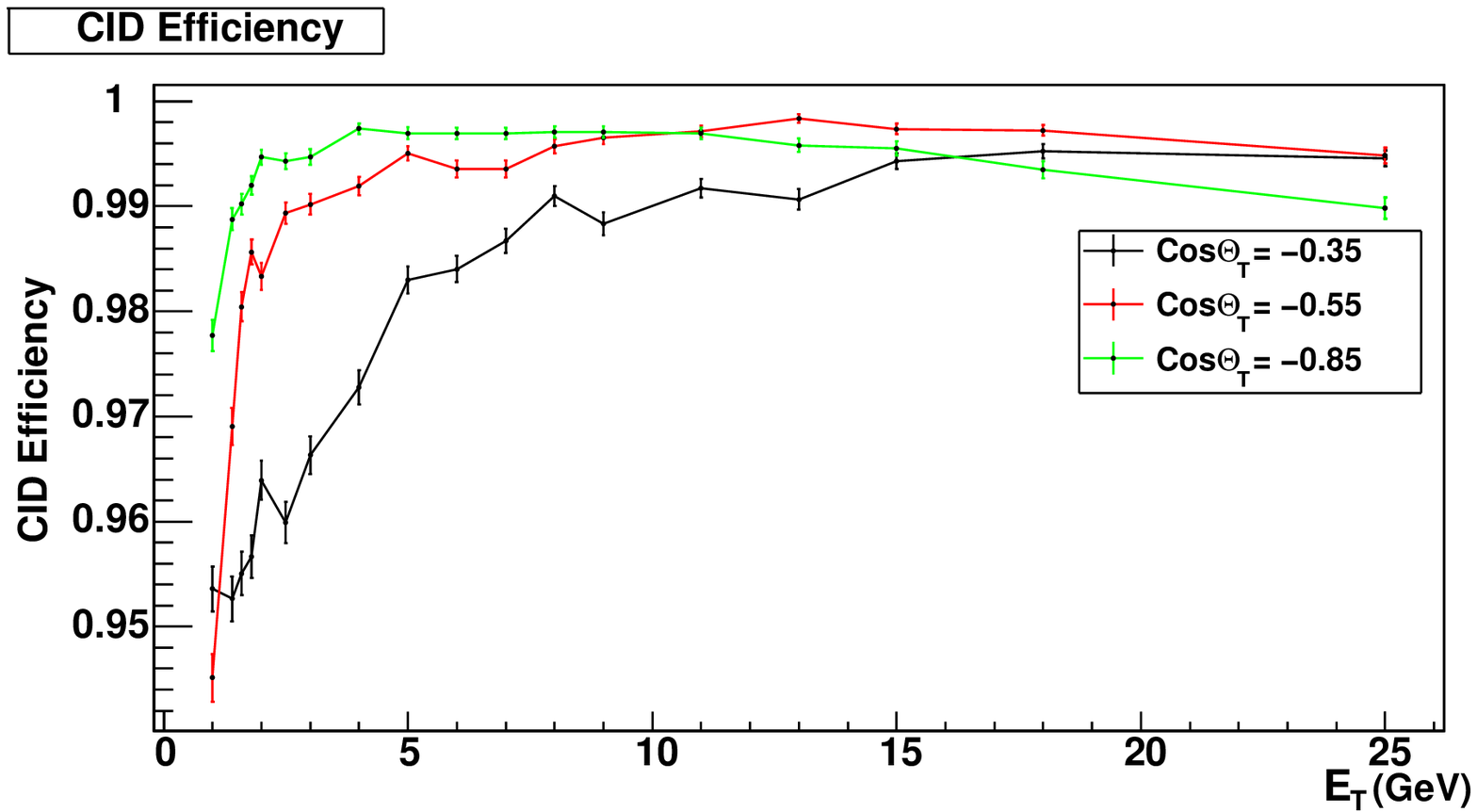}

\includegraphics[width=0.49\textwidth]{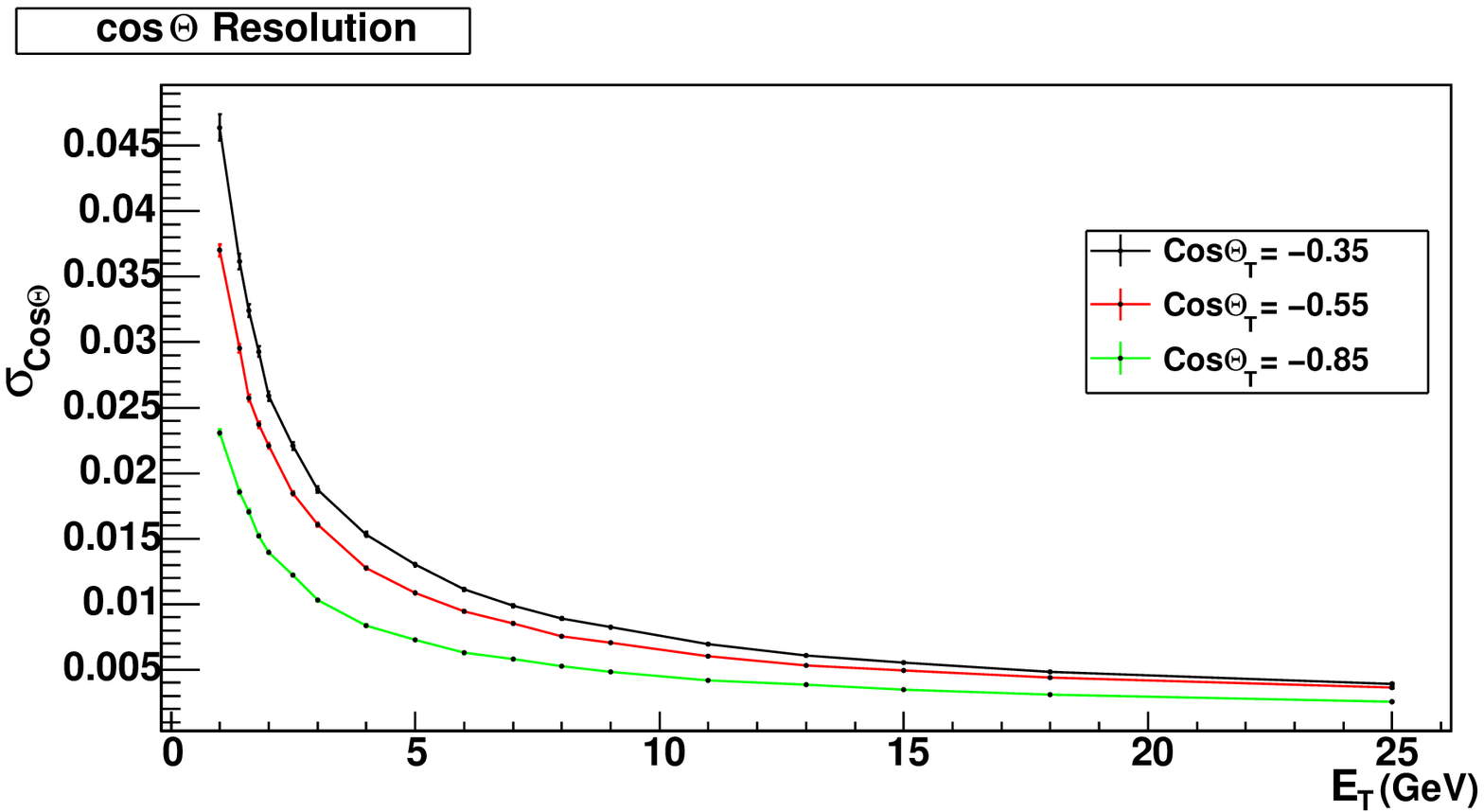}
\includegraphics[width=0.49\textwidth]{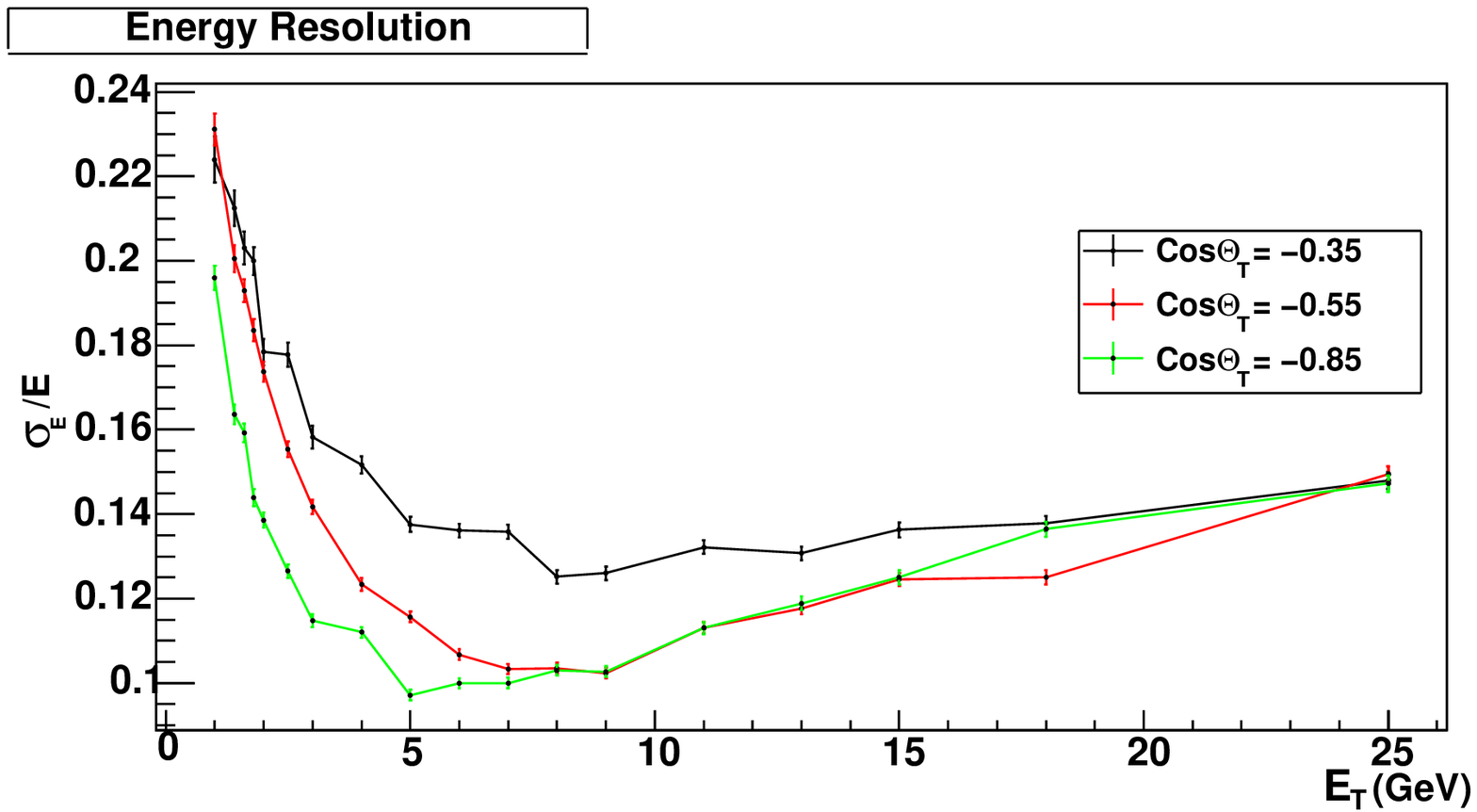}
\caption{The efficiencies and resolutions for muons in ICAL@INO as a function of 
the true muon energy. 
The top left panel shows the reconstruction efficiency of muons, the 
top right panel shows the charge identification efficiency. The bottom 
left panel shows the zenith angle resolution in $\cos\Theta$ while the bottom right 
panel shows the energy resolution of the muons. The lines in 
three colors correspond to three different benchmark zenith angles for the muons.
}
\label{fig:eff}
\end{figure}

\begin{figure}[t]
\centering
\includegraphics[width=0.49\textwidth]{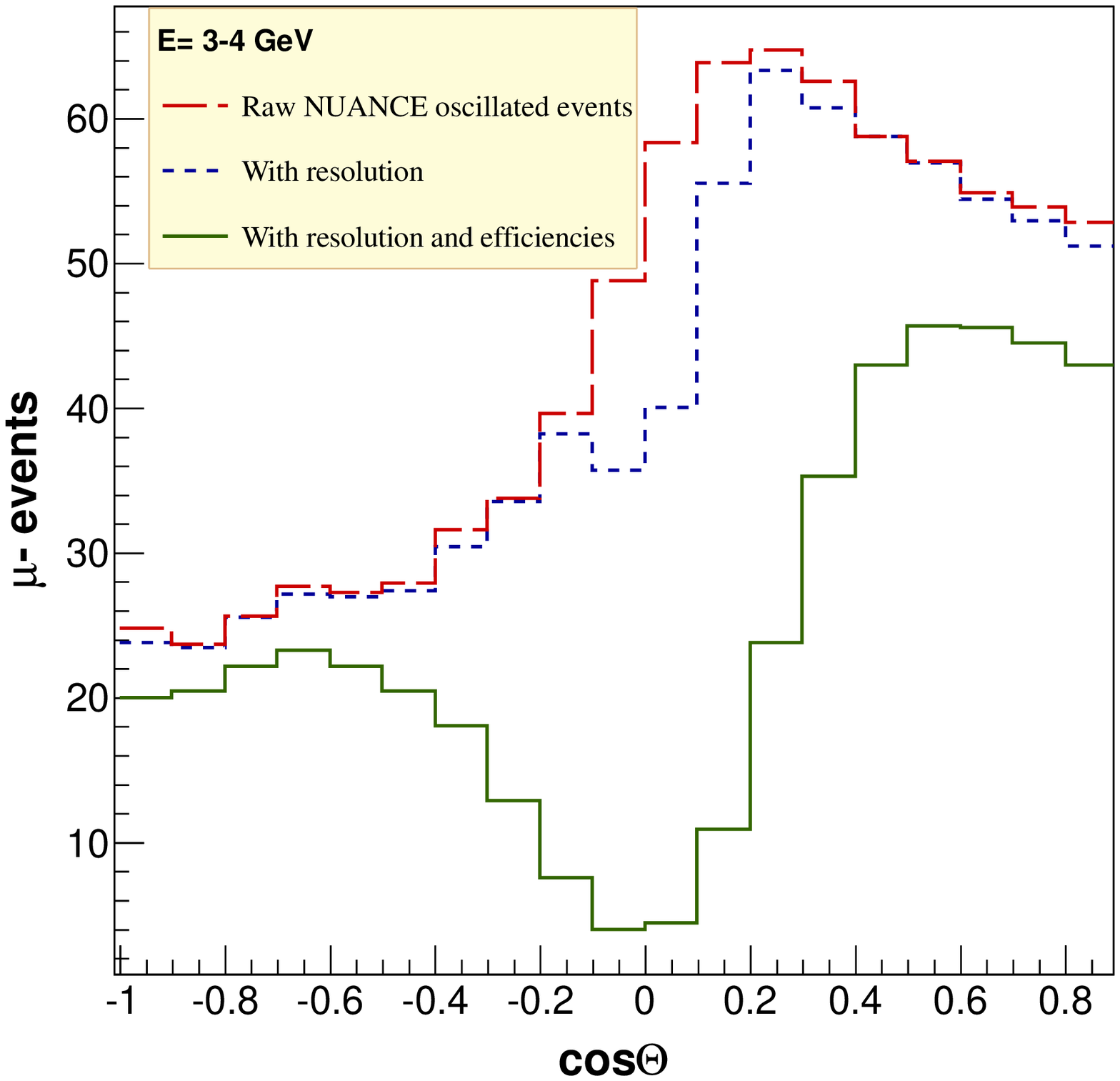}
\includegraphics[width=0.49\textwidth]{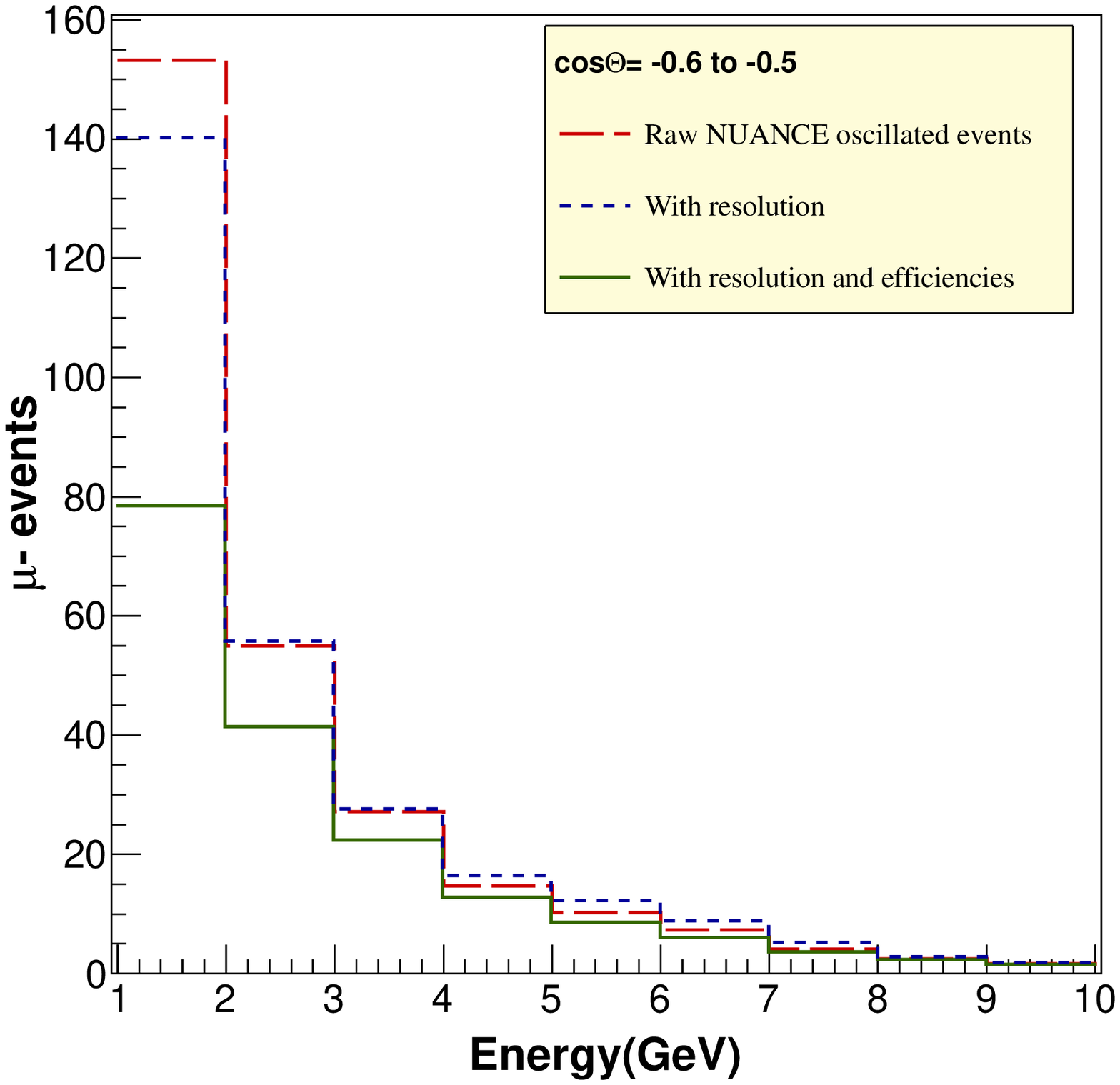}
\caption{The number of $\mu^-$ events for 10 years of running of ICAL@INO. The left panel 
shows the events in zenith angle bins for muon energy range $3-4$ GeV. 
The right panel shows the energy event spectrum in the zenith bin $-0.6$ to $-0.5$. 
The long-dashed 
(red) lines are the distribution of $\mu^-$ events for NUANCE events on which 
only the oscillations are imposed. The short-dashed (blue) lines give the zenith angle distribution 
when we fold the energy and angle resolution on the earlier event spectrum. The solid (green) lines are 
obtained when in addition to the resolutions we also fold in the efficiencies for muons 
in ICAL@INO.}
\label{fig:ev}
\end{figure}

For the atmospheric neutrino analysis presented in this paper, 
we have generated the unoscillated $\mu^-$ and $\mu^+$ events using the NUANCE 
event generator \cite{nuance} adapted for ICAL@INO\footnote{We use the 
Honda et. al atmospheric neutrino fluxes calculated for Kamioka \cite{honda}. While the 
atmospheric neutrino fluxes for Theni have been made available recently \cite{theni}, they  
are yet to be fully implemented in the ICAL@INO event generator.}.
In order to reduce the 
Monte Carlo fluctuations in the event sample, we generate events 
corresponding to $50\times 1000$ kton-years exposure, which 
corresponds to 1000 years of running of 
ICAL@INO. 
This event sample is finally normalized to a 
realistic number of years of running of ICAL@INO, when we 
statistically analyze the data in the end. Since it takes fairly 
long to run the NUANCE code to generate such a large event 
sample, running it over and over again for each set of oscillation parameter 
is practically impossible. Therefore, we run the event generator only once for no 
oscillations and thereafter impose the  reweighting
algorithm to generate the event sample for any set of oscillation 
parameters. This reweighting algorithm works according to the 
following prescription:

Every $\numu$ event 
given by the event generator is characterized by a certain muon  
energy and muon zenith angle, as well as a certain neutrino energy and neutrino 
zenith angle. 
For this neutrino energy and neutrino zenith angle, the 
probabilities $P_{\numu \numu}$ and $P_{\numu\nue}$ are 
calculated numerically for any given set of oscillation parameters. 
A random number $R$ between 0-1 is generated.
If $R < P_{\numu\nue}$ it is classified as a $\nue$ event. 
If $ R >  (P_{\numu\nue} + P_{\numu \numu})$, then we classify the 
event as a $\nutau$ event. 
If $P_{\numu\nue} \leq R \leq (P_{\numu\nue} + P_{\numu \numu})$, then 
it means that this event has come from an atmospheric $\numu$ which has 
survived as a $\numu$ and is hence selected as muon neutrino event.
Of course only the muon events are relevant for 
us, while the others are essentially discarded (in this work). 
Since we do this for a statistically large event sample, we get a 
$\numu$ ``survived" event spectrum that follows the survival probability to a high precision. 
One could also get muon events in the detector 
from oscillation of atmospheric $\nue$ into $\numu$. To find these events we 
generate events from NUANCE using atmospheric $\nue$ fluxes 
but $\numu$ charged current interactions in the detector, with the oscillation 
probability part of the code switched off. In order to get the oscillated 
muon event from in this sample 
we take a random number $S$ and use a similar procedure for 
classifying the events. That is, if $S < P_{\nue\numu}$, then the event 
is taken as an ``oscillated" $\numu$ event, which is the only part relevant for us in this work. 
The net number of $\mu^-$ events are obtained by adding the ``survived" and the 
``oscillated" $\numu$ events. 
The same exercise is performed for generating the 
$\mu^+$ events in the detector. 
We have checked that the final event spectrum we obtain 
with our method agrees remarkably well with the event spectrum obtained by 
passing the oscillated fluxes through the event generator. 

The data sample after incorporating oscillations is then distributed 
in very fine muon energy and zenith angle bins. This raw binned data is 
then folded with detector efficiencies and resolution functions 
to simulate the reconstructed muon events in ICAL. In our work we have 
used (i) the muon reconstruction efficiency, (ii) the muon charge identification 
efficiency, (iii) the muon zenith angle resolution and (iv) the muon 
energy resolution, which have been  
obtained by the INO collaboration \cite{inomuon}. The  
energy and zenith angle resolutions for the muons are in the form of a 
two-dimensional table. This means that the muon energy resolution  
is a function of both the muon energy as well as 
the muon zenith angle. Likewise the 
muon zenith angle resolution depends on both the muon energy as well as muon 
zenith angle. 
The muon detection efficiencies are also dependent on both the 
energy and zenith angle of the muon. Similarly the charge 
identification efficiency depends on both the energy and angle 
of the muon. In Fig. \ref{fig:eff} we show a snapshot of the efficiencies and resolutions 
for the $\mu^-$ events in ICAL@INO\footnote{
The detector response to muons used in this work comes from the first set of simulations done with the ICAL code. These simulations are on-going and are being improved. Hence the values of the resolution functions and efficiencies are likely to evolve along with the simulations. 
More details on this will appear shortly in a 
separate paper on the detector response to muons 
from the INO collaboration on the ICAL Geant based 
simulations \cite{inomuon}.}. 
They are shown as a function 
of the muon energy for three specific muon zenith angle bins of width 0.1 at 
$\cos\Theta_T = -0.85$, $-0.55$ and $-0.35$. The top-left panel shows the 
reconstruction efficiency of muons as a function of the muon energy. The top-right 
panel shows the charge identification efficiency. The bottom-left panel gives 
the muon zenith angle resolution, while the bottom-right panel shows the 
muon energy resolution. One can notice that there is a rather strong
dependence of all the four quantities on the muon energy as well 
zenith angle. The reconstruction efficiency, charge identification efficiency 
and the zenith angle resolution are seen to improve with muon energy. The 
muon energy resolution on the other hand shows a more 
complex behavior. While for energies $1-5$ GeV the $\sigma_E/E_T$ is 
seen to decrease as energy is increased, thereafter it increases with energy.
In the figure the detector performance is seen to be best for 
$\cos\Theta_T = -0.85$ and worst for $\cos\Theta_T = -0.35$. 
This is related to the geometry of the ICAL detector wherein there are 
horizontal slabs of iron and RPCs. As a result, the more horizontal muons 
travel longer in iron and hit a lesser number of RPCs. Therefore, the 
detector performs worse for more horizontal bins. In fact, for zenith 
bins $-0.1 \ltap \cos\Theta_T \ltap 0.1$ it becomes extremely difficult to reconstruct 
the muon tracks and hence for these range of zenith angles the reconstruction 
efficiency is effectively zero. The efficiencies and resolutions for 
$\mu^-$ and $\mu^+$ events in ICAL@INO have been obtained separately 
from simulations and 
are found to be 
similar \cite{inomuon}. We use the 
separate $\mu^-$ and $\mu^+$ efficiencies and resolutions in our simulations and results.

The number of $\mu^-$ events 
in the ${ij}^{th}$ bin after implementing the efficiencies and resolutions are given 
\be
 N_{ij}^{\prime th}(\mu^-) = {\cal N} \sum_k \sum_l K_i^k(E_T^k) \,\,
 M_j^l(\cos \Theta_T^l) 
 \bigg({\cal E}_{kl}{\cal C}_{kl}\,n_{kl}(\mu^-) + {\overline {\cal E}_{kl}}(1-\overline{{\cal C}_{kl}})\,n_{kl}(\mu^+)\bigg)
 \,,
 \label{eq:eventsth}
\ee
where ${\cal N}$ is the normalization required for a specific exposure in ICAL@INO, 
$E_T$ and $\cos\Theta_T$ are the 
true (kinetic) energy and true zenith angle of the muon, while $E$ and $\cos\Theta$ are the 
corresponding (kinetic) energy and zenith angle reconstructed from the 
observation of the muon track in the detector. 
The indices $i$ and $j$ correspond to the measured energy and zenith angle bins while 
$k$ and $l$ run over the true energy and zenith angle of the muons. 
The quantities $n_{kl}(\mu^-)$ and $n_{kl}(\mu^+)$ 
are the number of $\mu^-$ and $\mu^+$ events   
in the $k^{th}$ true energy and $l^{th}$ true zenith angle bin 
respectively, obtained by folding the raw events 
from NUANCE with the three-generation 
oscillation probabilities using the reweighting algorithm and subsequently 
binning the data as described earlier. The summation is over $k$ and $l$ where $k$ and $l$ scan all true 
energy and true zenith angle bins respectively. 
In Eq. (\ref{eq:eventsth}), 
${\cal E}_{kl}$ and $\overline{{\cal E}_{kl}}$ are the 
reconstruction efficiencies of $\mu^-$ and $\mu^+$ respectively, 
while ${\cal C}_{kl}$ and $\overline{{\cal C}_{kl}}$ are the 
corresponding charge identification efficiencies for $\mu^-$ and $\mu^+$ respectively,  
in the $k^{th}$ energy and 
$l^{th}$ zenith angle bin. 
Both the reconstruction 
efficiencies as well as charge identification efficiencies 
are functions of the true muon energy $E_T$ and true muon 
zenith angle $\cos\Theta_T$. The quantities $K_i^k$ and $M_j^l$ 
carry the information regarding the resolution functions of the detector and 
are seen to be 
\be
 K_i^k (E_T^k) = \int_{E_{L_i}}^{E_{H_i}} dE \,
 \frac{1}{\sqrt{2\pi} \sigma_{E}} 
  \exp \left( {- \frac{(E_T^k - E)^2}{2 \sigma_{E}^2} } \right) \,,
\label{eq:eresoln}
\ee
and
\be
M_j^l (\cos \Theta_T^l) = \int_{\cos \Theta_{L_j}}^{\cos \Theta_{H_j}} d\cos  \Theta\,
\frac{1}{\sqrt{2\pi} \sigma_{\cos\Theta} }
\exp \left (  - \frac{(\cos \Theta_T^l - \cos \Theta)^2}{2 \sigma_{\cos\Theta}^2} \right ) \,,
\label{eq:zenithresoln}
\ee
respectively. The resolution functions are seen to be Gaussian from ICAL simulations 
\cite{inomuon} above 1 GeV. 
The 
resolution functions $\sigma_{E}$ and $\sigma_{\cos\Theta}$ are 
obtained from ICAL simulations \cite{inomuon} and depend on both 
$E_T$ and $\cos\Theta_T$. A snapshot of these were shown in the lower panels of 
Fig. \ref{fig:eff}. 
An expression similar to Eq. (\ref{eq:eventsth}) 
can we written for the $\mu^+$ events $N_{ij}^{\prime th} (\mu^+)$.

In Fig. \ref{fig:ev} 
we show the $\mu^-$ event distribution expected in ICAL@INO 
with 10 years of exposure. In the left panel the events are shown for the energy 
bin $3-4$ GeV and 
in $\cos\Theta$ bins of width 0.1. The right panel shows events in the zenith 
range $\cos\Theta = -0.6$ to $-0.5$ and in energy bins of width 1 GeV. 
The red long-dashed lines shows the NUANCE events 
obtained after including the effect of oscillations according to the 
reweighting algorithm described above. 
The oscillation parameters used are given in Table \ref{tab:param},  with 
$\sa=0.5$, $\stch=0.1$ and assuming normal hierarchy. The blue short-dashed lines in the 
figure are obtained once the oscillated events (red long-dashed lines) in ICAL@INO are folded with the  
energy and zenith angle resolutions. A comparison of the red long-dashed and 
blue short-dashed lines in the right panel of the figure reveals that the effect of the energy resolution 
is to flatten the shape of the energy spectrum. We notice that the blue 
short-dashed line falls below the 
red long-dashed line for lower energy bins, while the trend is reversed for the 
the higher energy bins. On the other hand, 
the impact of the angle resolution is seen to be negligible 
for most of the zenith angle bins, as can be seen from from left panel of the figure. 
The reason for these features 
can be found in the size of the $1\sigma$ width of the resolution functions 
shown in Fig. \ref{fig:eff}. The energy resolution is seen to be $\sigma_E \simeq 0.15 E$ and 
hence for $E$ between $1-11$ GeV we do expect some spill-over 
between bins leading to a smearing 
of the energy spectrum. On the other hand, the bottom right panel of Fig. \ref{fig:eff} 
reveals that the $\cos\Theta$ resolution for most of the zenith angle bins 
are seen to be better than $\cos\Theta \sim 0.01-0.02$, while the zenith bins 
that we have used in the figure are $\Delta \cos\Theta=0.1$ in width. This is why the smearing 
due to angle resolution is essentially inconsequential for this choice of zenith angle 
binning. In section \ref{sec:bin} we will discuss in detail the impact of bin size on the 
mass hierarchy sensitivity of ICAL@INO. From a careful and detailed analysis, 
we will choose an optimal bin size both in energy as well as zenith angle. 

The green solid lines in Fig. \ref{fig:ev} show the realistic events in ICAL@INO where 
we have taken into account oscillations, detector resolutions as well as reconstruction 
and charge identification efficiencies. Meaning, these lines are obtained by imposing 
the reconstruction 
and charge identification efficiencies on the red short-dashed lines. 
The left panel shows that once the detector efficiencies are folded, 
the number of events go to almost 
zero for the horizontal bins. This happens because of difficulty in reconstructing 
the muon tracks along the nearly horizontal directions as discussed before. 

\begin{figure}[t]
\centering
\includegraphics[width=0.3\textwidth]{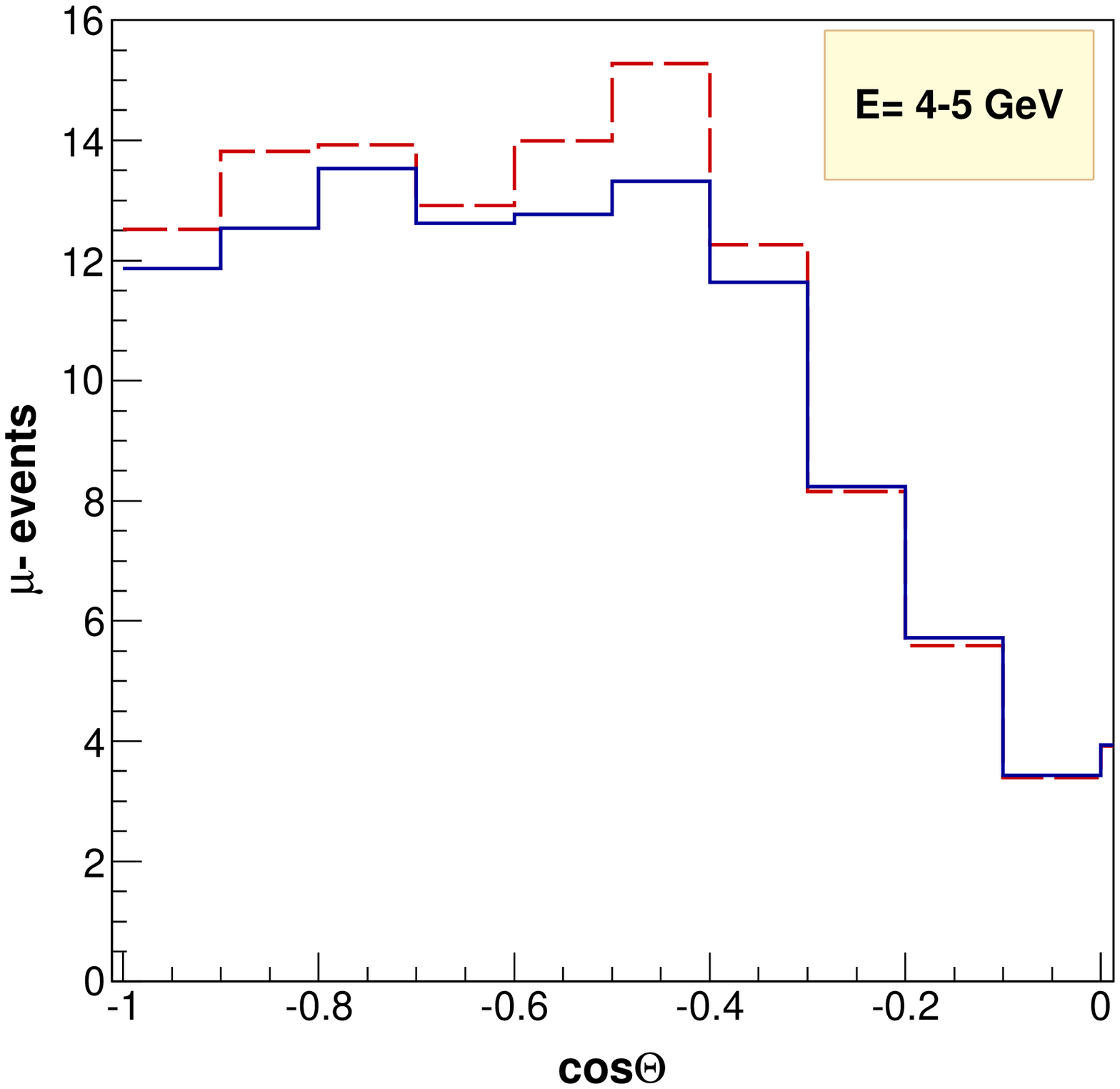}
\includegraphics[width=0.3\textwidth]{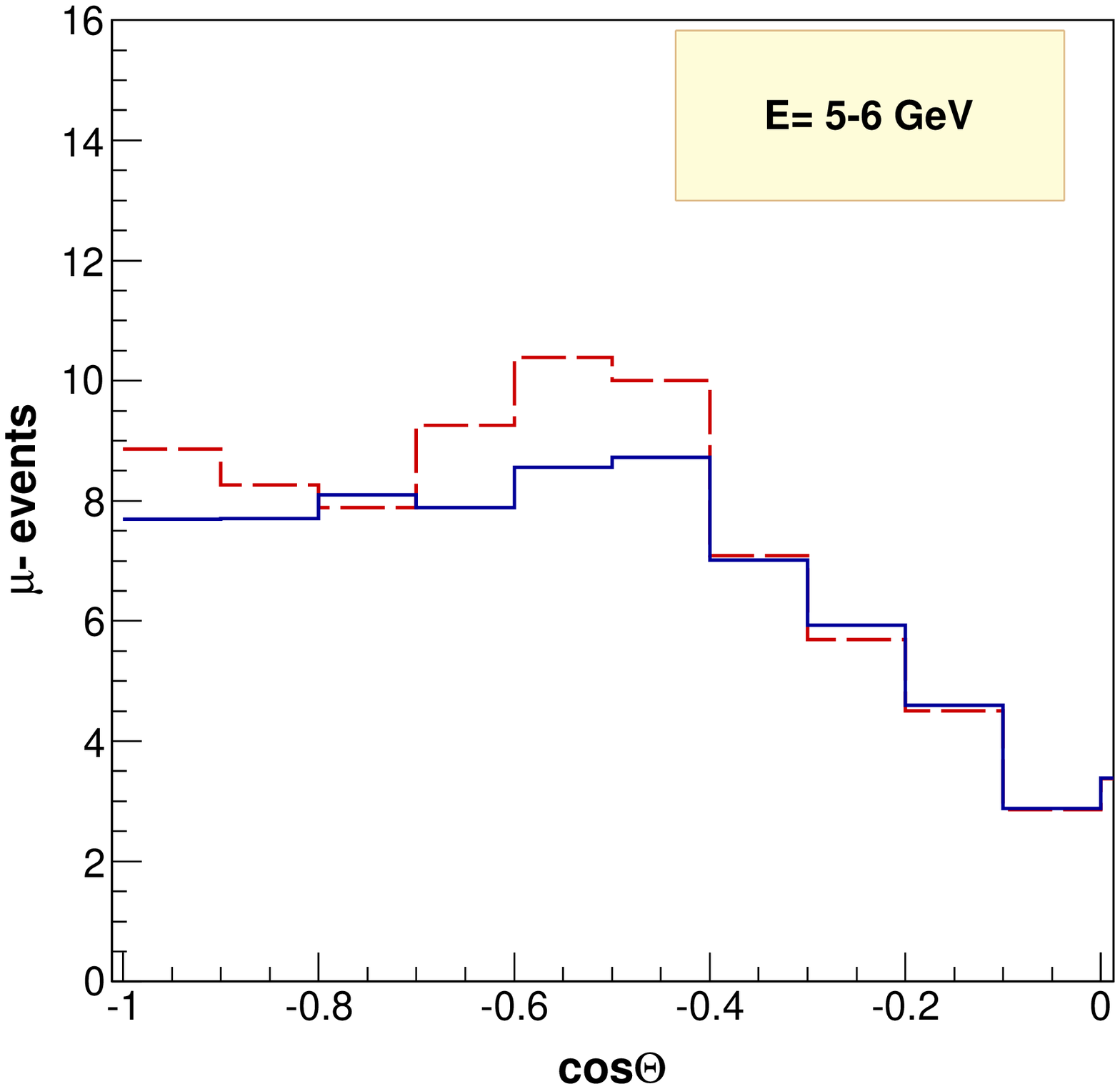}

\includegraphics[width=0.3\textwidth]{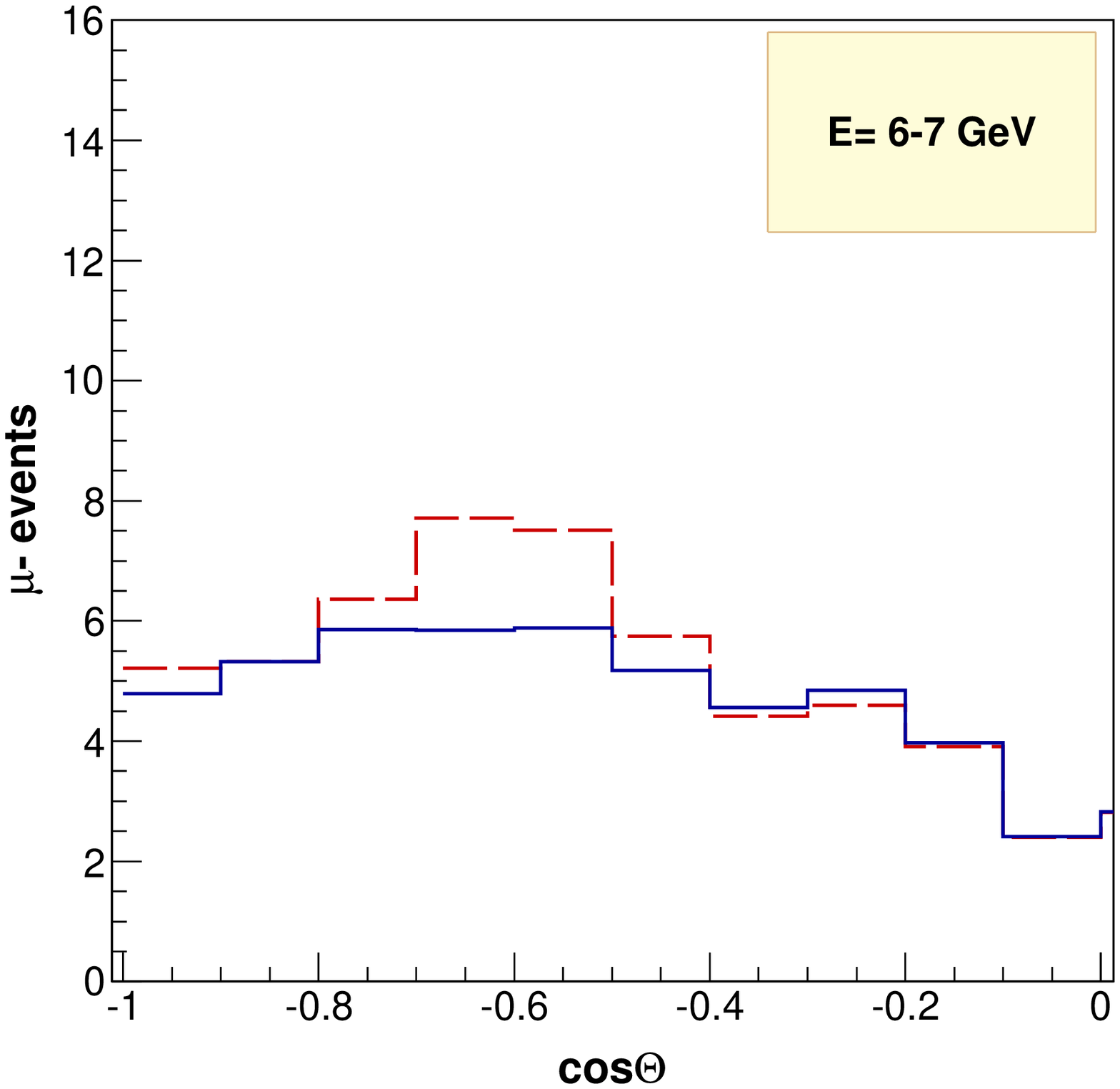}
\includegraphics[width=0.3\textwidth]{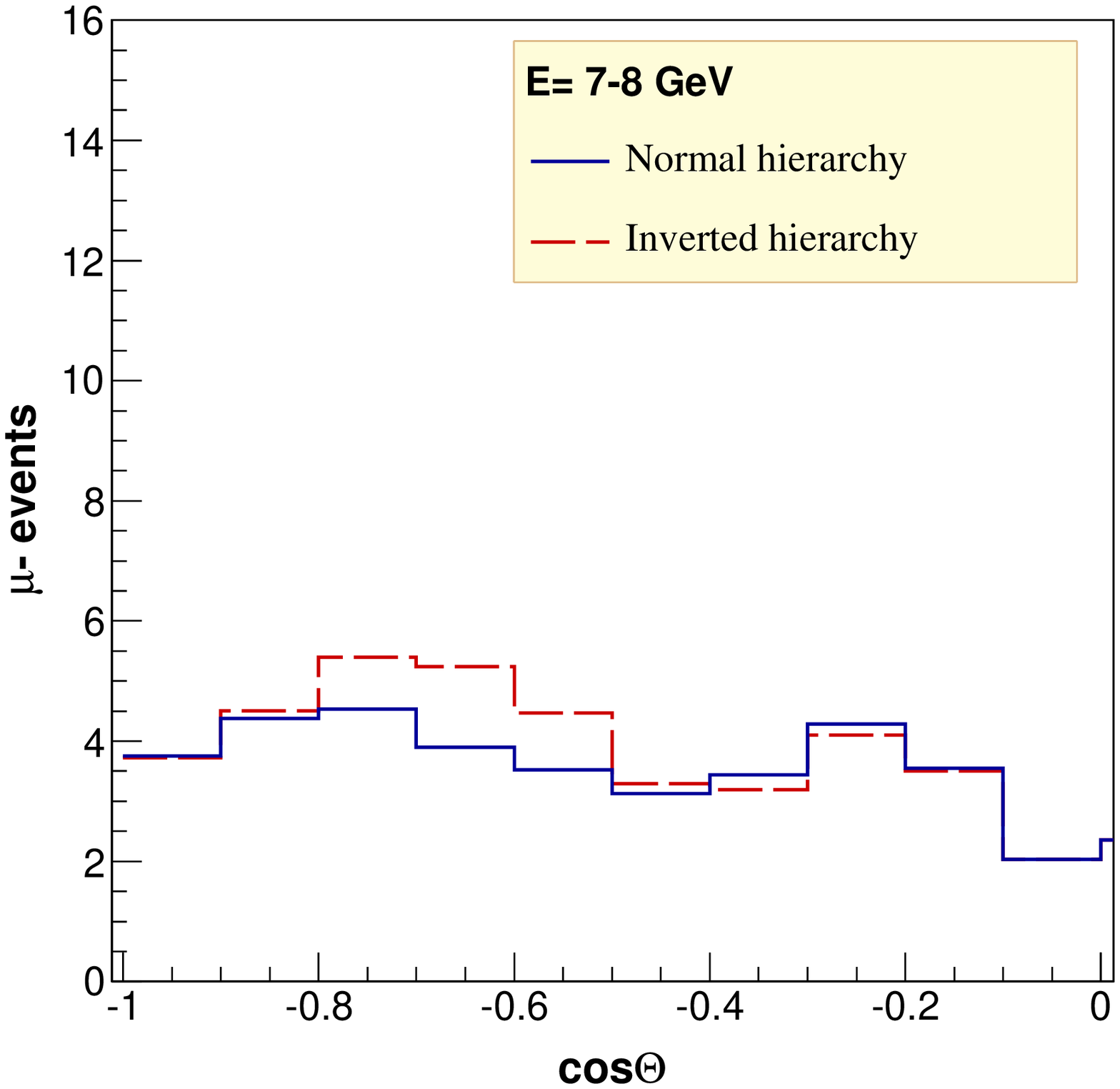}
\caption{
ICAL@INO $\mu^-$ 
event spectrum in zenith angle bins for four specific energy bins of 
$E=4-5$ GeV (top left panel), $E=5-6$ GeV (top right panel), $E=6-7$ GeV (bottom left panel), 
and $E=7-8$ GeV (bottom right panel) and for 10 years exposure.
The solid blue lines 
correspond to $N_i^{\prime th} (\mu^-)$ for normal hierarchy 
while red-dashed lines are for inverted hierarchy. 
}
\label{fig:evhier}
\end{figure}

Fig. \ref{fig:evhier} shows the ICAL@INO $\mu^-$ 
event spectrum in zenith angle bins for four specific muon energy bins of 
$E=4-5$ GeV (top left panel), $E=5-6$ GeV (top right panel), $E=6-7$ GeV (bottom left panel), 
and $E=7-8$ GeV (bottom right panel) and for 10 years exposure.
The solid blue lines 
correspond to $N_i^{\prime th} (\mu^-)$ for the normal hierarchy 
while red-dashed lines are for the inverted hierarchy. Events were generated 
at the benchmark oscillation point given in Table \ref{tab:param}, with $\sa=0.5$ and 
$\stch=0.1$. 
We can see that 
for normal hierarchy there are earth matter effects in the $\mu^-$ 
channel leading to suppression of the event spectrum. 
The extent of the suppression is seen to depend on both 
energy as well as zenith angle bin of the $\mu^-$. 
The good energy and angle resolution of the detector 
is crucial here for fine enough binning of the events to extract maximum 
effect of the earth matter effects. 
For inverted 
hierarchy there are no earth matter effects in the $\mu^-$ events. 
On the other hand, for $\mu^+$ earth matter effects appear 
for inverted hierarchy and are absent for the normal hierarchy. 
This is why charge separation is so crucial for mass hierarchy 
determination. If the $\mu^-$ and $\mu^+$ events were to be added 
one would lose sensitivity to earth matter effects and hence to 
neutrino mass hierarchy. ICAL@INO will have excellent 
charge identification capabilities and zenith angle resolution 
function as well as good energy resolution.

\section{The Statistical Analysis}

In what follows, we generate the data at the benchmark true values for oscillation 
parameters given in Table \ref{tab:param} and assuming a certain neutrino mass hierarchy. 
We then  fit this simulated data with the wrong mass hierarchy to check the statistical 
significance with which this wrong hierarchy can be disfavored. 
For doing this statistical test we define a $\chi^2$ for ICAL@INO data as
\be
\chi^2_{ino}(\mu^-)={\min_{\{\xi_k\}}} \sum_{i=1}^{N_i} \sum_{j=1}^{N_j}  \left
[2\left(N_{ij}^{th}(\mu^-)-N_{ij}^{ex}(\mu^-)\right) 
+2N_{ij}^{ex}(\mu^-)ln\left(\frac{N_{ij}^{ex}(\mu^-)}{N_{ij}^{th}(\mu^-)}\right)\right] + 
\displaystyle\sum\limits_{k=1}^l \xi_k^{2}
\,,
\label{eq:chisqino}
\ee
where 
\be
N_{ij}^{th}(\mu^-)=N_{ij}^{\prime{th}}(\mu^-)\left(1+\displaystyle\sum\limits_{k=1}^l \pi_{ij}^{k}{\xi_k}\right) + 
{\cal O}(\xi_k^2)
\,.
\label{eq:evth}
\ee
We have assumed Poissonian distribution for the errors in this definition of $\chi^2$. 
The reason is that for the higher energy bins $E\simeq 5-10$ GeV where we expect to 
see the hierarchy sensitivity, the number of events fall sharply (cf. Fig. \ref{fig:evhier})
and for small exposure times these bins could have very few events per bin. 
Since ICAL@INO will have separate data in $\mu^-$ and $\mu^+$, we 
calculate this $\chi^2_{ino}(\mu^-)$ and $\chi^2_{ino}(\mu^+)$ separately 
for the $\mu^-$ sample and the $\mu^+$ 
sample respectively and then add the two to get the $\chi^2_{ino}$ as
\be
\chi^2_{ino} = \chi^2_{ino}(\mu^-) + \chi^2_{ino}(\mu^+)
\,.
\label{eq:chiino}
\ee
In the above equations,  $N_{ij}^{ex}(\mu^-)$ and $N_{ij}^{ex}(\mu^+)$ 
are the observed number of $\mu^-$ and $\mu^+$ events respectively in the $i^{th}$ 
energy and $j^{th}$ zenith angle bin 
and 
$N_{ij}^{\prime{th}}(\mu^-)$ and $N_{ij}^{\prime{th}}(\mu^+)$ are the 
corresponding theoretically predicted event 
spectrum given by Eq. (\ref{eq:eventsth}). 
This predicted event spectrum could shift due to the systematic  uncertainties 
and this shifted spectrum $N_{ij}^{th}$ is given by Eq. (\ref{eq:evth}). In the above 
 $\pi_{ij}^k$ is the $k^{th}$ systematic uncertainty in the ${ij}^{th}$ bin and 
$\xi_k$ is the pull variable corresponding to the uncertainty $\pi^{k}$. 
The $\chi^2_{ino}$ is minimized over the full set of pull variables $\{\xi_k\}$. 
The index 
$i$ runs from 1 to the total number of energy bins $N_{i}$ and $j$ runs from 1 to 
total number of zenith angle bins $N_j$. In our analysis we 
have considered the muon energy range 1 GeV to 11 GeV, while the  
zenith angle range in $\cos \Theta$ is taken from $-1$ to $+1$. 
We will discuss in some detail the impact of binning on the mass hierarchy sensitivity 
of ICAL@INO in section \ref{sec:bin}.
The index $k$ in Eqs. (\ref{eq:chisqino}) and 
(\ref{eq:evth}) runs from 1 to $l$, where $l$ is the total number of systematic uncertainties.
We have included the following five systematic uncertainties in our analysis. An overall 
flux normalization error of 20\% is taken. A 10\% error is taken 
on the overall normalization of the cross-section. A 5\% uncertainty on 
the zenith angle dependence of the fluxes 
is included. 
An energy dependent ``tilt factor" is considered according to the 
following prescription. The event spectrum is calculated with the predicted 
atmospheric neutrino fluxes
and then with the flux spectrum shifted according to
\be
\Phi_\delta (E) = \Phi_0 (E)\bigg (\frac{E}{E_0}\bigg)^\delta \simeq \Phi_0 (E) \bigg ( 1 + \delta 
\, \ln \frac{E}{E_0} \bigg )
\,,
\ee
where $E_0=2$ GeV and $\delta$ is the $1\sigma$ systematic 
error which we have taken as 5\%. The difference 
between the predicted events rates for the two cases is then included in the 
statistical analysis. 
Finally, an over all 5\% systematic 
uncertainty is included.

For the analysis of the data from the current accelerator and reactor experiments we have used 
the individual $\chi^2_i$  for each experiment as defined in GLoBES \cite{globes}.

\section{Mass Hierarchy Sensitivity -- Effect of Binning \label{sec:bin}}

\begin{figure}
\centering
\includegraphics[width=0.49\textwidth]{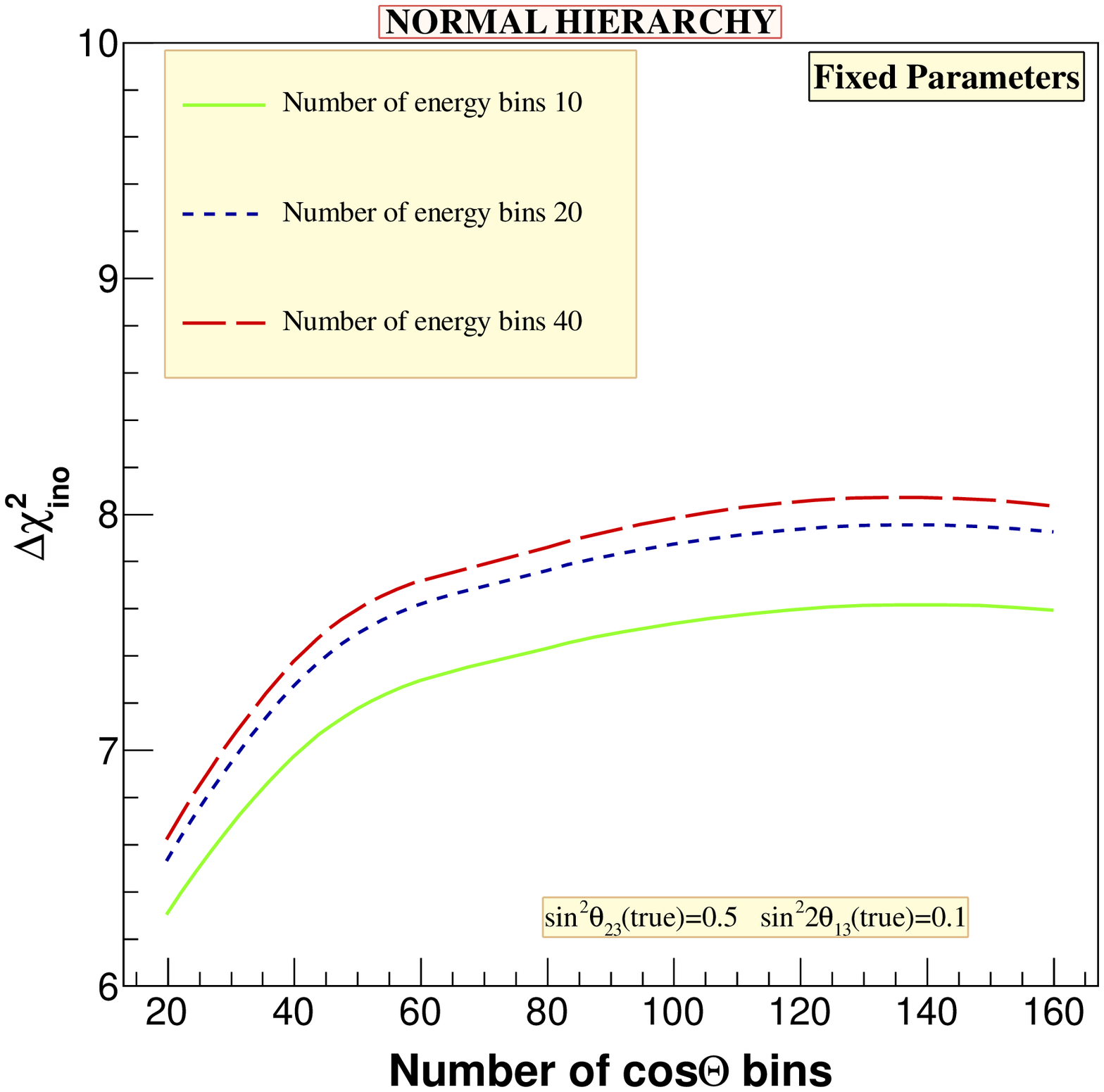}
\includegraphics[width=0.49\textwidth]{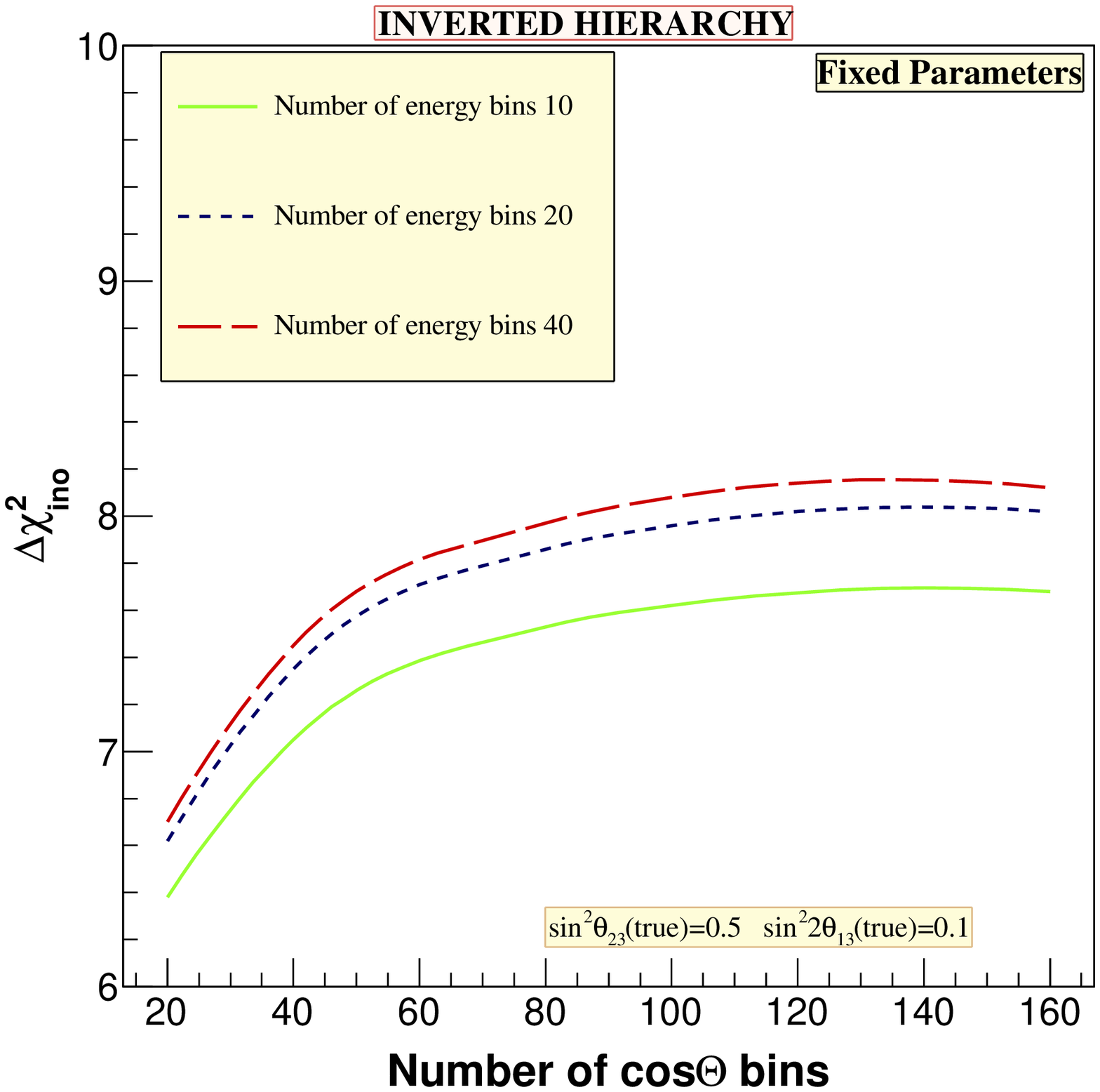}
\caption{Effect of binning on the mass hierarchy reach of ICAL@INO with 10 years data.}
\label{fig:hierbin}
\end{figure} 

The earth matter effect in atmospheric neutrinos is known to fluctuate rapidly 
with energy as well as zenith angle.
Therefore, 
one would like to observe the difference in energy and zenith angle spectrum 
between the $\mu^-$ and $\mu^+$ as accurately as possible. 
Since averaging of data in energy and/or zenith angle bins is expected to 
reduce the sensitivity of the experiment to earth matter effect and hence to the 
neutrino mass hierarchy, one would ideally like to perform an 
unbinned likelihood analysis of the ICAL@INO data. 
This will obviously be the approach once the real data of ICAL@INO is available. 
However, at this stage we can only perform a binned $\chi^2$ analysis of the 
simulated ICAL@INO data. In the following, we begin by first analyzing the 
effect of bin size on the mass hierarchy sensitivity of ICAL@INO and choose 
the optimum bin sizes in energy and zenith angle of the muons.

In Fig. \ref{fig:hierbin} we show the effect of bin size on the mass hierarchy sensitivity of ICAL@INO. 
The ICAL@INO data used in this figure 
corresponds to 10 years of running of the experiment and is generated 
at the benchmark values of the oscillation given in Table \ref{tab:param} 
and for $\stch=0.1$ and $\sa=0.5$. Since our main objective here is to look at the impact of 
bin size on the mass hierarchy sensitivity, we do not bring in the complication of marginalization 
over the oscillation parameters and keep all oscillation parameters fixed at 
their true values in the fit. We show the $\chi^2_{ino}$ obtained as a function of the 
number of bins in $\cos\Theta$, where the zenith angle $\cos\Theta$ ranges between 
$-1$ and $+1$. The left panel is for normal hierarchy taken as true while the right 
panel is for true inverted hierarchy. The three lines in each of the panels are for three 
different choice for the number of energy bins. The solid green lines are for 10 energy bins, 
the dashed blue lines for 20 energy bins while the long-dashed red lines correspond to 
40 energy bins. The range of measured muon energy considered in each case is 
between 1 GeV and 11 GeV. We notice that the $\chi^2_{ino}$ increases as the number of 
$\cos\Theta$ bins is increased from 20 and eventually flattens out beyond 80. Likewise 
the sensitivity is seen to increase as we increase the number of energy bins. However, 
there is no substantial gain beyond the case for 20 energy bins. 
This trend agrees well with the ICAL@INO resolutions obtained in energy and $\cos\Theta$, a 
snapshot of which is reported in Fig. \ref{fig:eff}.  The $\cos\Theta$ resolution 
$\sigma_{\cos\Theta} \sim 0.025$ which corresponds to 80 bins, while 
energy resolution at $E_\mu \simeq 5$ GeV is seen to be $\sigma_E \sim 0.1$ which 
is compatible with energy bin size of 0.5 GeV with 20 energy bins. 
Therefore, in the rest of the this paper, we work with 20 energy and 80 $\cos\Theta$ bins.

\section{Mass Hierarchy Sensitivity -- Main Results}

\begin{figure}
\centering
\includegraphics[width=0.48\textwidth,height=8.2cm]{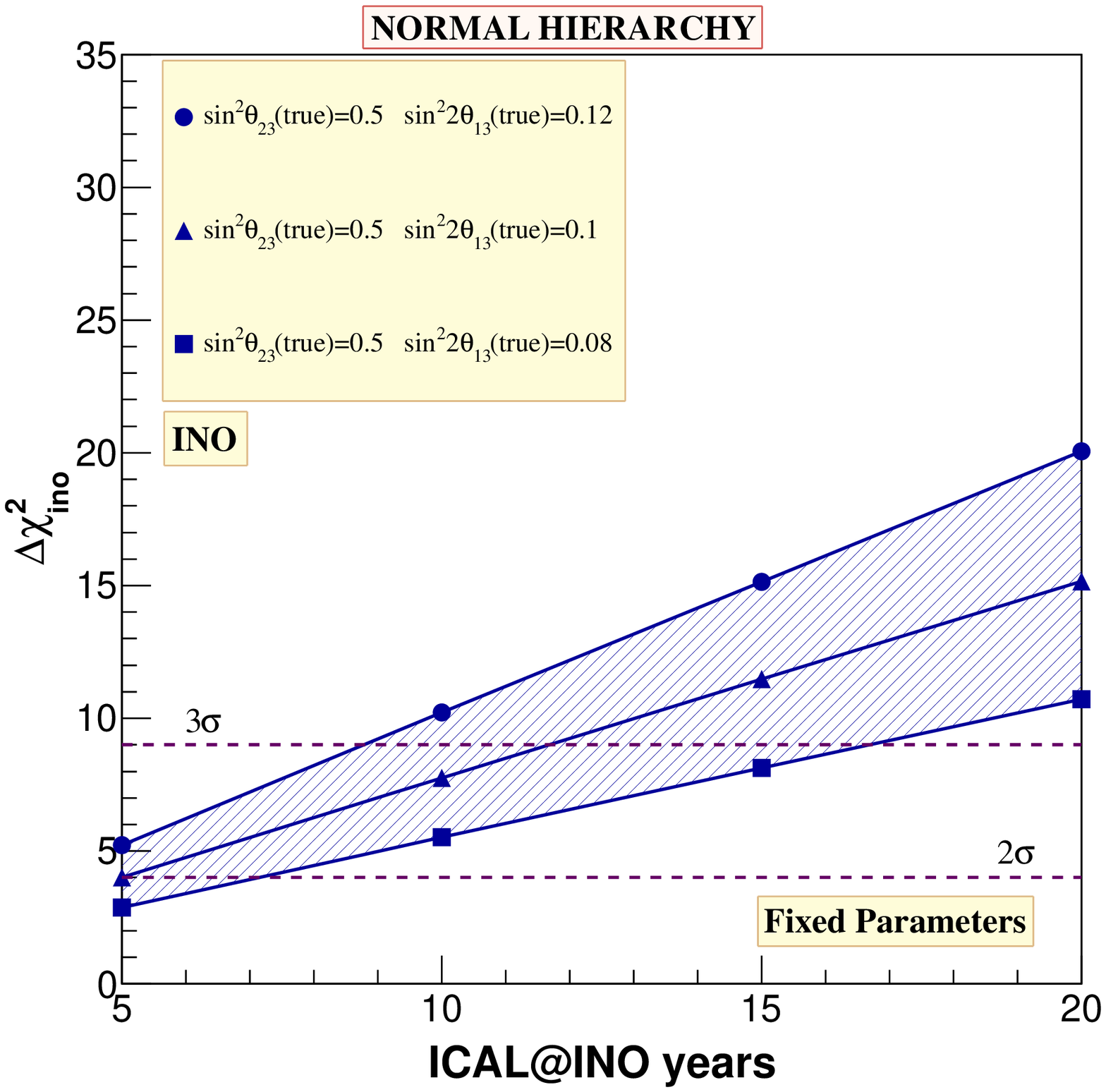}
\includegraphics[width=0.48\textwidth]{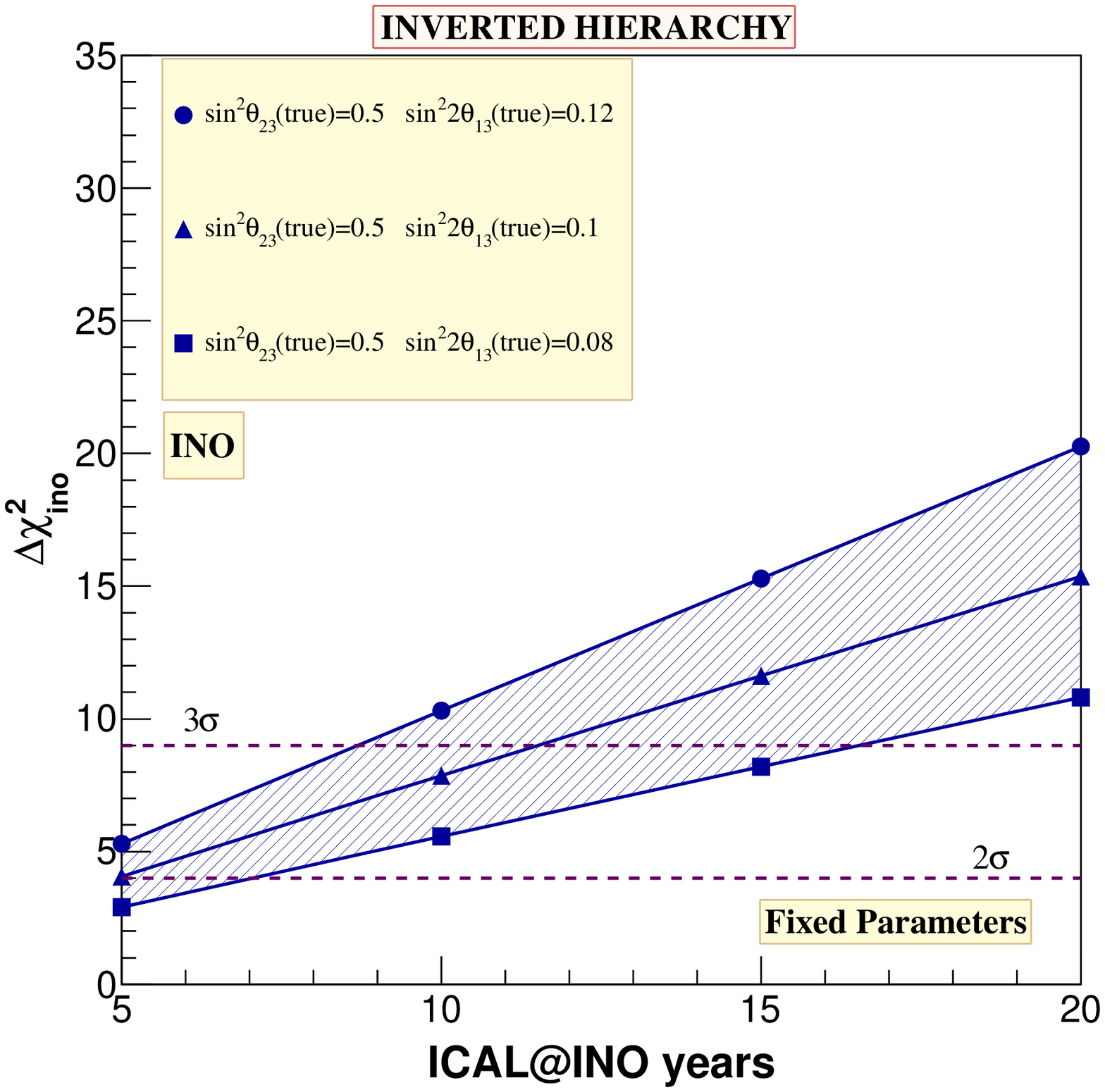}
\caption{Left panel shows the $\Delta \chi^2_{ino}$ for the wrong hierarchy when normal hierarchy is taken 
to be true, while the right panel shows the corresponding reach when inverted hierarchy is taken as true. 
The three lines are for three different values of $\stcht=0.08$, 0.1 and 0.12 as shown in the legend 
box, while $\sat=0.5$ for all cases. We take only ICAL@INO data into the analysis and in the fit 
keep all oscillation parameters fixed at their benchmark true values.}
\label{fig:hierfixed}
\end{figure}

\begin{figure}
\centering
\includegraphics[width=0.48\textwidth]{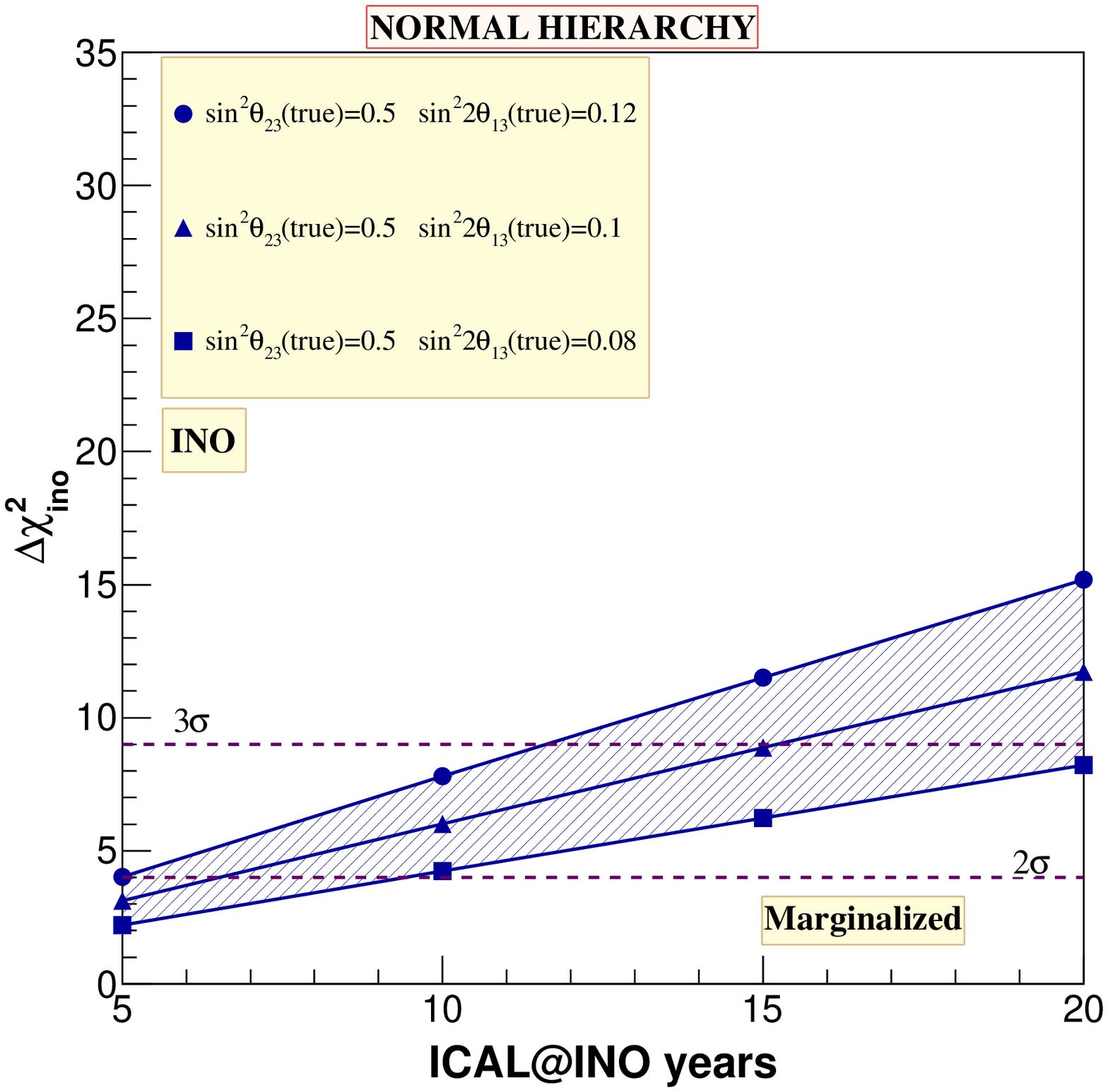}
\includegraphics[width=0.48\textwidth]{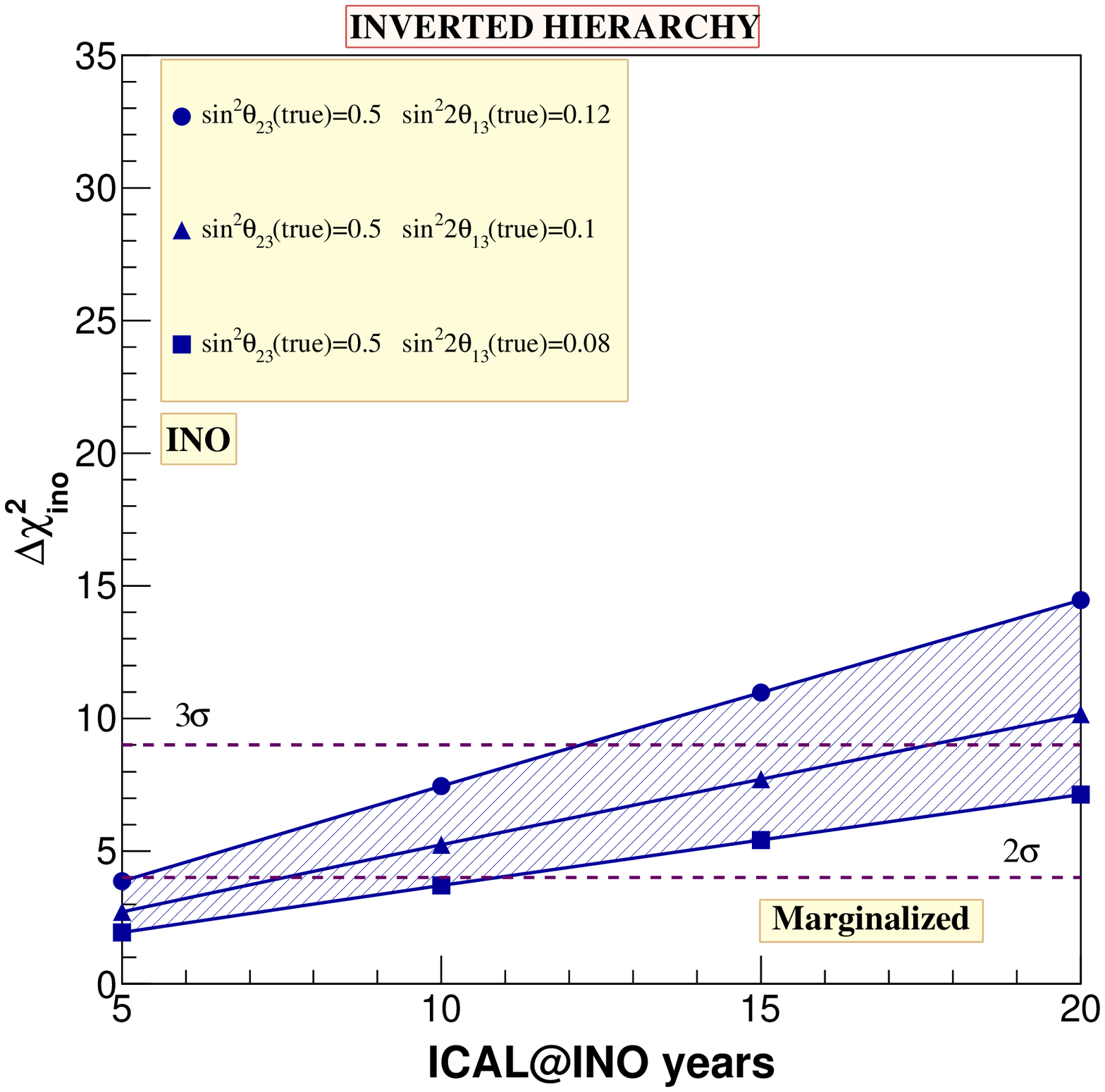}
\caption{Same as Fig. \ref{fig:hierfixed} but here oscillation parameters $|\meff|$, $\sa$ and 
$\stch$ are allowed to vary freely within their current $3\sigma$ ranges given in Table \ref{tab:param}. 
}
\label{fig:hiermarg}
\end{figure}

\begin{figure}[t]
\centering
\includegraphics[width=0.48\textwidth]{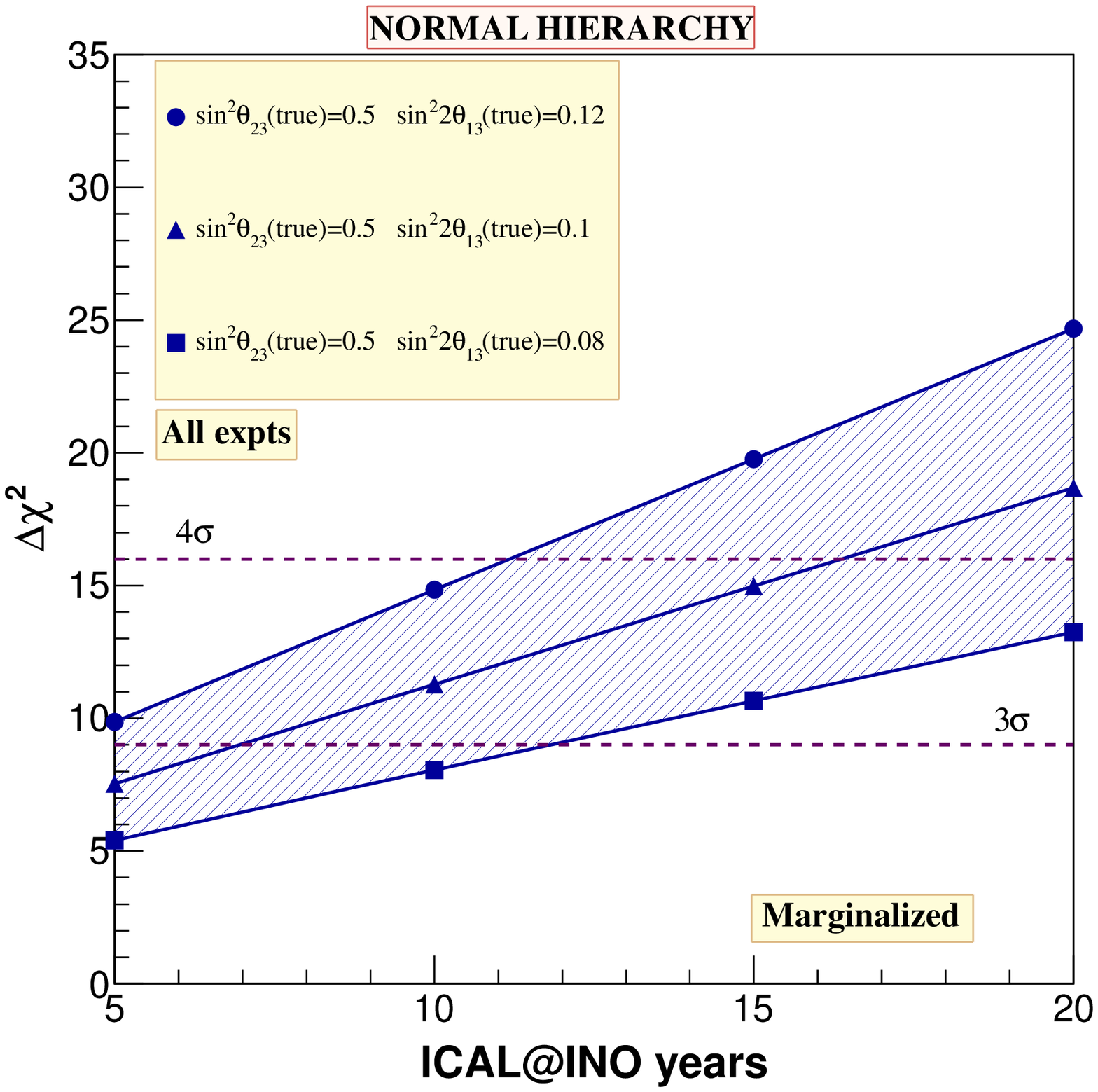}
\includegraphics[width=0.48\textwidth]{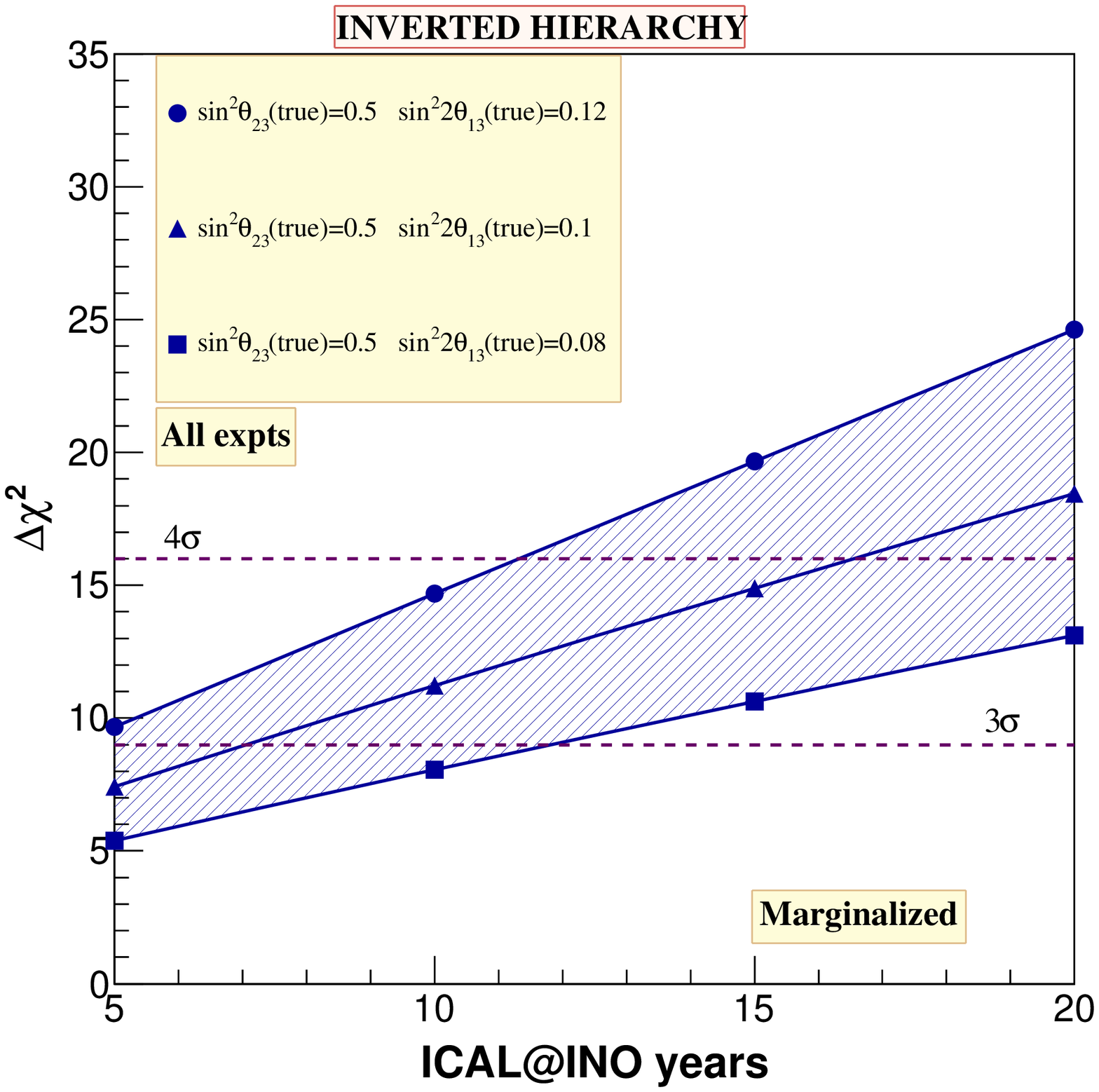}
\caption{The $\Delta \chi^2$ for the wrong hierarchy 
obtained from a combined analysis of all experiments including ICAL@INO as well 
NO$\nu$A, T2K, Double Chooz, RENO and Daya Bay experiments. The left panel is for 
normal hierarchy taken as true in the data while the right panel is for inverted hierarchy as true.
The three lines are for three different values of $\stcht=0.08$, 0.1 and 0.12 as shown in the legend 
box, while $\sat=0.5$ for all cases. In the fit we allow all parameters to vary within their 
$3\sigma$ ranges as shown in Table \ref{tab:param}.
}
\label{fig:hiersens}
\end{figure}

In Fig. \ref{fig:hierfixed} we show the discovery potential 
of ICAL@INO alone for the neutrino mass hierarchy, 
as a function of the number of years of running of the experiment.
The data is generated for the values of the oscillation parameters 
given in Table \ref{tab:param} and for $\sat=0.5$. The three lines correspond to 
the different values of $\stcht$ shown in the 
legend boxes in the figure, which we have chosen around the best-fit and $2\sigma$ 
range of the current best-fit. 
The left-hand panel is for true normal hierarchy while 
the right-hand panel is for true inverted hierarchy. These plots show the 
sensitivity reach of ICAL@INO when all oscillation parameters are kept 
fixed in the fit at the values at which the data was generated. For $\stcht$ around the 
current best-fit of 0.1, 
we can note from the these plots that with 5 years of 
ICAL@INO data alone, we will have a $2.0\sigma$ ($2.0\sigma$) signal for the wrong 
hierarchy if normal (inverted) hierarchy is true. After 10 years of ICAL@INO 
data, this will improve to 
$2.8\sigma$ ($2.8\sigma$) signal for the wrong 
hierarchy if normal (inverted) hierarchy is true. 
The sensitivity obviously increases with the true value of 
$\stcht$. The $\Delta \chi^2$ is seen to increase almost linearly with 
exposure. This is not hard to understand as the hierarchy sensitivity comes 
from the difference in the number of events between normal and 
inverted hierarchies due to earth matter effects. Since this is a small difference,
the relevant statistics in this measurement is small. As a result the 
mass hierarchy analysis is statistics dominated and one can see from 
Eq. (\ref{eq:chiino}) that in the statistics dominated regime the $\Delta \chi^2_{ino}$ 
increases linearly with exposure.

The hierarchy sensitivity quoted above are for fixed values of the oscillation parameters. 
This effectively means that the values of all oscillation parameters are known with 
infinite precision. Since this is not the case, the sensitivity will go down once we 
take into account the uncertainty in the value of the oscillation parameters. 
The oscillation parameters which affect the mass hierarchy sensitivity of ICAL@INO the most 
are $|\meff|$, $\sa$ and $\stch$. 
In Fig. \ref{fig:hiermarg} 
we show the mass hierarchy sensitivity reach of ICAL@INO 
with full marginalization over the 
oscillation parameters $|\meff|$, $\sa$ and $\stch$, 
meaning these oscillation parameters are allowed to 
vary freely in the fit within their {\it current} $3\sigma$ ranges, and the minimum of 
the $\chi^2$ taken from that. 
The CP phase $\delta_{CP}$ 
does not significantly 
impact the ICAL@INO mass hierarchy sensitivity. This will be discussed in some 
detail later. Therefore, we keep $\delta_{CP}$ fixed at 0 in the fit. 
The parameters $\ms$ and $\sin^2\theta_{12}$ also do not affect  
$\chi^2_{ino}$ and hence are kept fixed at their true values given in Table \ref{tab:param}. 
From the figure we see that for full marginalization within the {\it current} 
$3\sigma$ allowed range for $|\meff|$, $\sa$ and $\stch$, 
the sensitivity reach of ICAL@INO with 10 (5) years data 
would drop to $2.5\sigma$ ($1.8\sigma$) for $\sat=0.5$ and $\stcht=0.1$, for 
true normal hierarchy. The impact for the inverted hierarchy case is seen to be more. 
However, this is not a fair way of assessing the sensitivity reach of the 
experiment since all values of the oscillation parameters are not 
allowed with equal C.L. by the current data. This implies that when 
they deviate from their current 
best-fit value in the fit, they should pick up a $\chi^2$ from the data of the experiment(s) 
which constrains them. Therefore, one should do a global fit taking all 
relevant data into account to find the correct estimate of the 
reach of combined neutrino data to the neutrino mass hierarchy. One way to 
take this into account is by introducing priors on the parameters and adding the 
additional $\chi^2_{prior}$ in the fit, analogous to what 
we had explained in Eq. (\ref{eq:prior}) and 
the related discussion for the priors on the solar parameters. 
Moreover, 
all oscillation parameters are expected to be measured with much better 
precision by the on-going and up-coming neutrino experiments. 
In fact, by the time 
ICAL@INO is operational,  all of the current accelerator-based and reactor 
experiments would have completed their scheduled run and hence 
we expect that by then significant 
improvements in the allowed ranges of the oscillation parameters would have been made.
In particular, we expect improvement 
in the values of $\stch$, $|\meff|$ and $\sta$ from the data 
coming from current accelerator and reactor experiments. 
A projected combined sensitivity analysis of these experiments shows that  the 
$1\sigma$ uncertainties on the values of $\stch$, $|\meff|$ and $\sta$
are expected to go down to 0.1\%, 2\% and 0.65\%, respectively \cite{tenyrs09}. 
Since marginalization over these parameters makes a difference to 
$\chi^2_{ino}$ for the wrong hierarchy (cf. Figs. \ref{fig:hierfixed} and \ref{fig:hiermarg}), 
better measurement on them from the current 
experiments will therefore improve the mass hierarchy 
sensitivity reach of ICAL@INO. 
As mentioned before, one could incorporate this information into 
the analysis by including ``priors" on these parameters. 
The sensitivity reach of ICAL@INO with 
projected priors on $|\meff|$ and $\sta$ keeping other parameters fixed can be 
found in \cite{choubeynu2012}. In the plot presented in \cite{choubeynu2012} 
$1\sigma$ prior of 2\% on $|\meff|$ and 0.65\% on 
$\sta$ was assumed. 
Note however, that in that analysis only two systematic 
uncertainties were included in the fit, an overall flux normalization uncertainty of 20\% and an 
overall cross-section uncertainty of 10\%. We will discuss the impact of 
systematic uncertainties again 
in section \ref{sec:syst}. 
In this work we improve the analysis by performing 
a complete global fit of the atmospheric neutrino 
data at ICAL@INO 
combined with all relevant data which would be available at that time, {\it viz.}, 
data from the full run of the T2K, NO$\nu$A, Double Chooz, Daya Bay, and 
RENO experiments. The combined sensitivity 
to the neutrino mass hierarchy as a function of number of years of run of the 
ICAL@INO atmospheric neutrino experiment is shown in Fig. \ref{fig:hiersens}. 
For each set of oscillation parameters, the joint $\chi^2$ from all experiments is given by 
\be
\chi^2 = \chi^2_{ino} + \sum_i \chi^2_i
\ee
where $\sum_i \chi^2_i$ is the contribution from 
the accelerator and reactor experiments and $i$ runs over 
T2K, NO$\nu$A, Double Chooz, Daya Bay, and 
RENO experiments. This 
joint $\chi^2$ is computed and  
marginalized over all oscillation parameters. 
The minimized joint $\Delta \chi^2$ is shown in Fig. \ref{fig:hiersens}.  We reiterate that the 
$x-$axis in this figure shows the number of years of running of ICAL@INO only, 
while for all other experiments we have considered their complete 
run as planned in their letter of intent and/or Detailed Project Report, 
as mentioned in section \ref{sec:lbl}. 
The left panel of the figure shows the sensitivity reach if normal hierarchy is 
true while the right panel shows the reach when the inverted hierarchy is the 
true hierarchy. As in Figs. \ref{fig:hierfixed} and \ref{fig:hiermarg} we generate the 
data at the values of the oscillation parameters given in Table \ref{tab:param} and 
with $\sat=0.5$ and three different values of $\stcht = 0.08$, 0.1, 
and 0.12. The figure shows that inclusion of the accelerator and reactor data 
increases the sensitivity such that 
with just 5 years of ICAL@INO data one 
would have more than $2\sigma$ evidence for the neutrino 
mass hierarchy even if $\stcht=0.08$. For the current 
best-fit of $\stcht=0.1$ we would rule out the 
wrong hierarchy at $2.7\sigma$ while for larger $\stcht=0.12$ mass hierarchy 
could be determined with about $3.1\sigma$ C.L.. With 10 years of ICAL@INO 
data the sensitivity would improve to $2.8\sigma$ for $\stcht=0.08$, $3.4\sigma$ for 
$\stcht=0.1$ and $3.9\sigma$ for $\stcht=0.12$. 

The inclusion of the accelerator and reactor experiments 
into the analysis improves the 
sensitivity reach to the neutrino mass hierarchy in the following two ways. 
Firstly, inclusion of these data sets into the analysis effectively restricts the 
allowed ranges of oscillation parameters $|\meff|$, $\sa$ and 
$\stch$ such that the statistical significance of the mass hierarchy determination 
from ICAL@INO alone goes up to what we were getting in Fig. \ref{fig:hierfixed} 
for fixed values of the oscillation parameters. In addition, we also get a 
contribution to the mass hierarchy sensitivity from the accelerator and reactor 
experiments themselves. 
 
 \begin{table}[h]
\begin{center}
\begin{tabular}{|c|c|c|c|c|c|c|c|c|}
\hline
&&&&&&&\\[-2.5mm]
Expts & NOvA & T2K & DB& RENO & DC & INO & ALL \\
&&&&&&&\\[-2.5mm]
\hline
$\Delta \chi^2_{\meff}$ & 2.59 & 0.26 & 0.53 & 0.12 & 0.02 & 7.76 & 11.28 \\\hline
$\Delta \chi^2_{\ma}$ & 2.49 & 0.31 & 0.63 & 0.14 & 0.02 & 7.95 & 11.53\\
\hline
\end{tabular}
\caption{\label{tab:contr}
Contribution to the $\Delta \chi^2$ towards the wrong mass hierarchy 
at the global best-fit from the 
individual experiments. The normal hierarchy was taken as true and data was 
generated at the benchmark true values of the oscillation parameters given in 
Table \ref{tab:param} with $\sat=0.5$ and $\stcht=0.1$. The first row ($\Delta \chi^2_{\meff}$) 
shows the individual contributions to the statistical significance when $\meff$ is 
used for defining the normal ($\meff>0$) and inverted ($\meff < 0$) mass 
hierarchy, while the lower row 
($\Delta \chi^2_{\ma}$) gives the corresponding contributions when $\ma>0$ is 
taken as normal hierarchy and $\ma <0$  as inverted hierarchy. Effect of 
choice of the definition for normal and inverted hierarchy will be discussed in 
the Appendix.
}
\end{center}
\end{table}

We show in Table \ref{tab:contr} the separate contributions from the individual 
experiments to the statistical significance for the mass hierarchy sensitivity 
from the global fit. The data is generated for normal hierarchy and the 
benchmark true values of the oscillation parameters given in 
Table \ref{tab:param} with $\sat=0.5$ and $\stcht=0.1$. The oscillation 
parameters are allowed to vary freely in the fit and the minimum 
global $\chi^2$ selected. The Table \ref{tab:contr} 
shows the individual $\Delta \chi^2$ contributions from each experiment 
at this global best-fit point for the oscillation parameters, as well as the 
combined $\Delta \chi^2$. The global best-fit for the inverted hierarchy 
corresponds to $\ms=7.5\times 10^{-5}$ 
eV$^2$, $\sss=0.31$, 
$\sa=0.5$, $\stch=0.1$, $\meff=-2.4\times 10^{-3}$ eV$^2$ and 
$\delta_{CP}=252^\circ$. 
Note that (cf. Eq. (\ref{eq:meff})) since 
$\ma$ depends on the value of $\dcp$, the change of $\dcp$ in the fit gives 
$|\Delta m_{31}^2| = 2.34\times 10^{-3}$ eV$^2$ at the global best-fit for the inverted 
hierarchy ($\meff <0$). 
However, the 
data was generated at $\meff=+2.4\times 10^{-3}$ eV$^2$, which for 
$\dcpt=0^\circ$ gives 
$|\Delta m_{31}^2| = 2.44\times 10^{-3}$ eV$^2$.
 
Table \ref{tab:contr} shows that it is mainly ICAL@INO and NO$\nu$A which contribute 
to the $\Delta \chi^2$ since reactor experiments have no sensitivity to the neutrino mass 
hierarchy and the baseline for T2K is far too short to allow for any significant 
earth matter effect in the signal. The NO$\nu$A experiment on the other hand 
has a baseline of 810 km and a higher energy neutrino beam. This gives the experiment 
sizable earth matter effects which in turn brings in sensitivity to the neutrino mass 
hierarchy. The small $\Delta \chi^2$ contribution from the reactor experiments 
comes from the fact that the best-fit value of the oscillation parameters, and 
in particular $|\ma|$ which controls the spectral shape of these experiments 
is slightly different at the global best-fit for inverted hierarchy, as discussed above.
The T2K experiment on the other hand returns a small contribution since the best-fit 
$\delta_{CP}$ is different from $\dcpt=0$ taken in data. 


The results in this section including those shown in Table \ref{tab:contr} 
are for $\delta_{CP}$(true)=0. 
However, 
the accelerator-based long baseline experiments are sensitive to 
the $\numu \to \nue$ oscillation channel. 
The size of this $P_{\numu \nue}$ oscillation probability and  
the resultant sensitivity depends crucially on the CP phase $\delta_{CP}$. 
Therefore, the contribution from NOvA to the statistical significance with 
which we can determine the neutrino mass hierarchy will depend crucially 
on the value of $\delta_{CP}$(true). 
We will discuss this in some detail in section \ref{sec:delta}.

\section{Impact of systematic uncertainties}
\label{sec:syst}

\begin{figure}
\centering
\includegraphics[width=0.48\textwidth,angle=0]{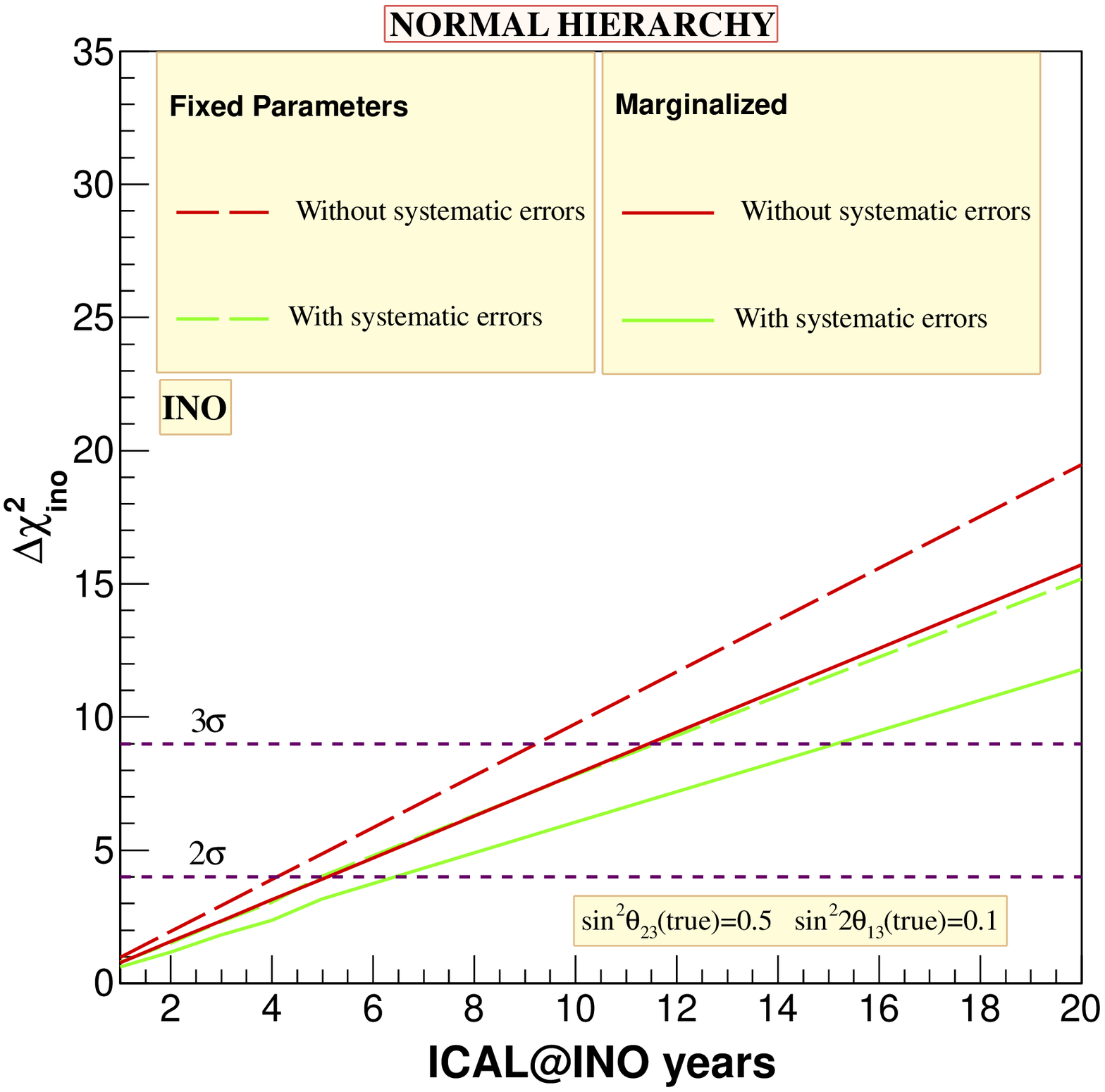}
\includegraphics[width=0.48\textwidth,angle=0]{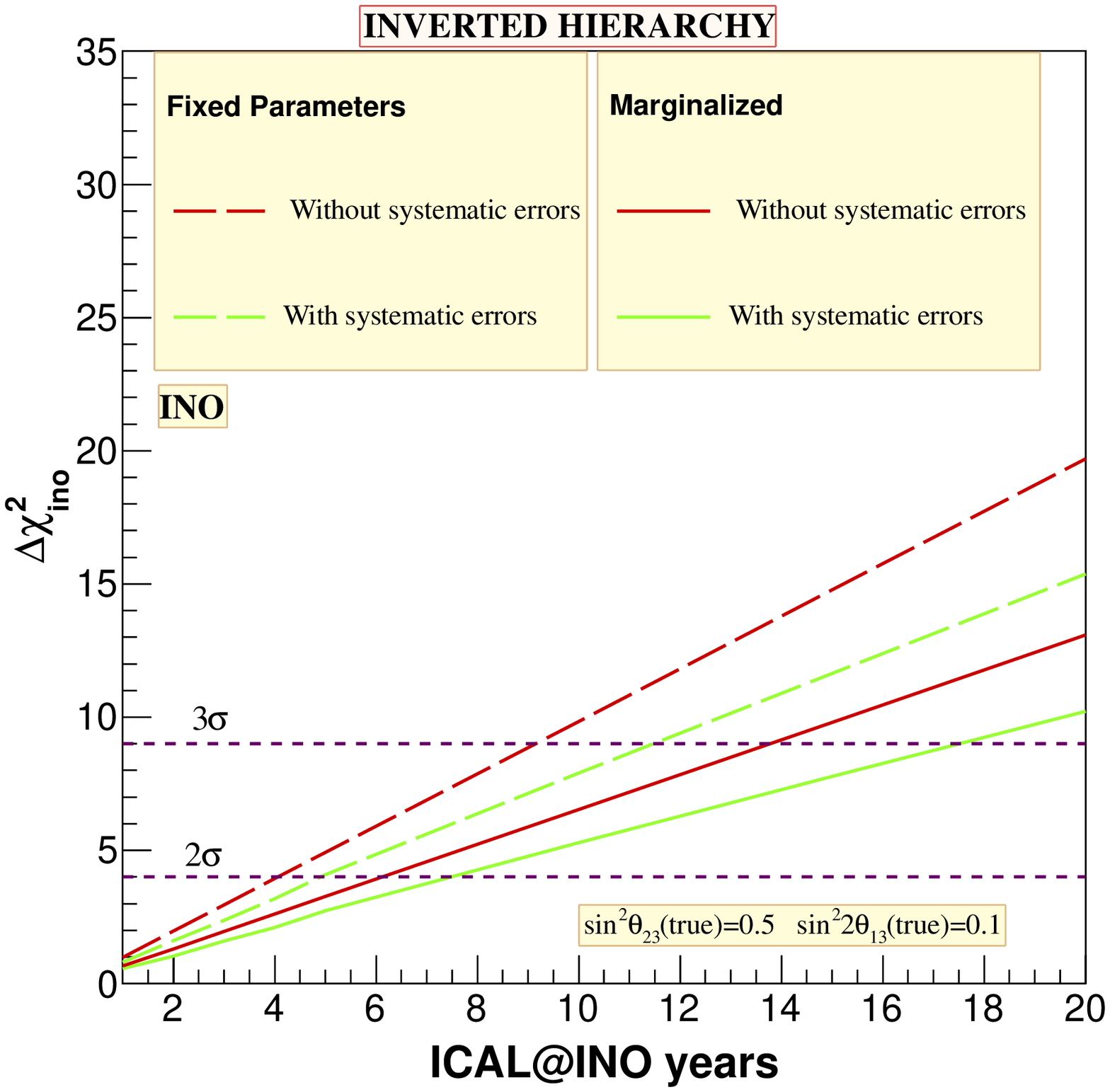}
\caption{The impact of systematic uncertainties on mass hierarchy sensitivity. The 
red lines are obtained without taking systematic uncertainties in the ICAL@INO analysis, 
while the green lines are obtained when systematic uncertainties are included. Long-dashed lines 
are for fixed parameters in theory as in data, while solid lines are obtained by marginalizing over 
$|\meff|$, $\sa$ and $\stch$.}
\label{fig:hiersyst}
\end{figure}

The atmospheric neutrino fluxes have large systematic uncertainties. 
In order to study the impact of these systematic uncertainties on the 
projected reach of ICAL@INO to the neutrino mass hierarchy, we show 
in Fig. \ref{fig:hiersyst} the mass hierarchy sensitivity with and without 
systematic uncertainties in 
the ICAL@INO analysis. The $\Delta \chi^2$ is 
shown as a function of the 
number of years of exposure of the experiment. The data was 
generated at the benchmark oscillation point. 
The 
red lines are obtained without taking systematic uncertainties in the ICAL@INO analysis, 
while the green lines are obtained when systematic uncertainties are included. The long-dashed lines 
are for fixed parameters in theory as in the data, while the solid lines are obtained by marginalizing over 
$|\meff|$, $\sa$ and $\stch$. The left panel is for true normal hierarchy while the right panel 
is for true inverted hierarchy. The effect of taking systematic uncertainties is to 
reduce the statistical significance of the analysis. We have checked that of the 
five systematic uncertainties, the uncertainty on overall normalization of the fluxes and 
the cross-section normalization uncertainty have minimal impact on the final results. 
The reason for that can be understood from the fact that the atmospheric neutrinos come 
from all zenith angles and over a wide range of energies. The 
overall normalization uncertainty is the same for all bins, while 
the mass hierarchy dependent earth matter effects, are important only in certain zenith 
angle bin and certain range of energies. Therefore, the 
effect of the overall normalization 
errors get cancelled between different bins.
On the other hand, the tilt error could be used to 
modify the energy spectrum 
of the muons in the fit and the zenith angle error allows changes to the zenith angle 
distribution. Therefore, these errors do not cancel between the different bins and 
can dilute the significance of the data. In particular, we have checked that 
the effect of the zenith angle dependent systematic error on the atmospheric neutrino fluxes 
has the maximum effect on the lowering of the $\Delta \chi^2$ for the mass hierarchy 
sensitivity.

\section{Impact of $\sa$(true)}
\label{sec:sa}

\begin{figure}
\centering
\includegraphics[width=0.48\textwidth]{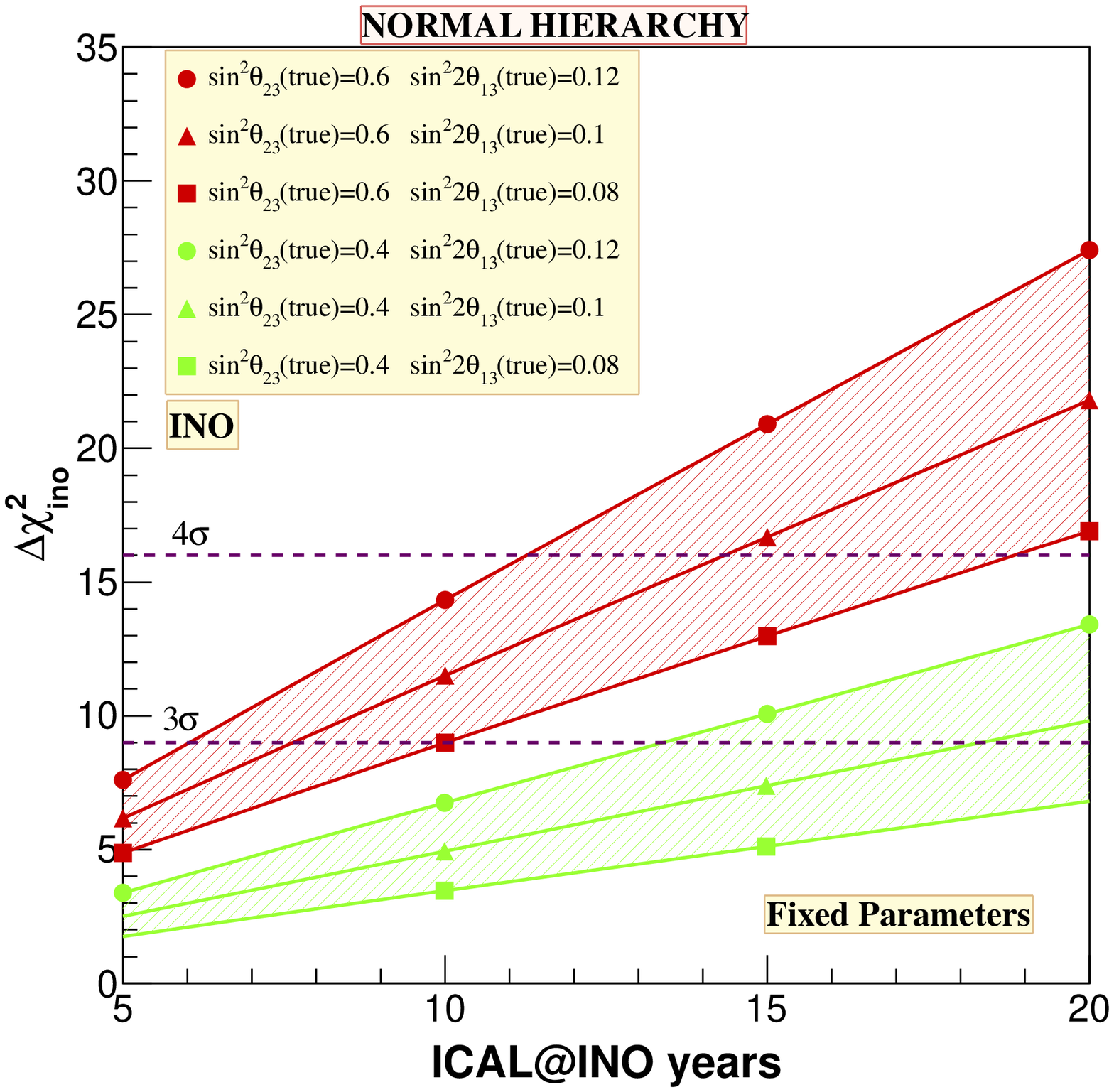}
\includegraphics[width=0.48\textwidth]{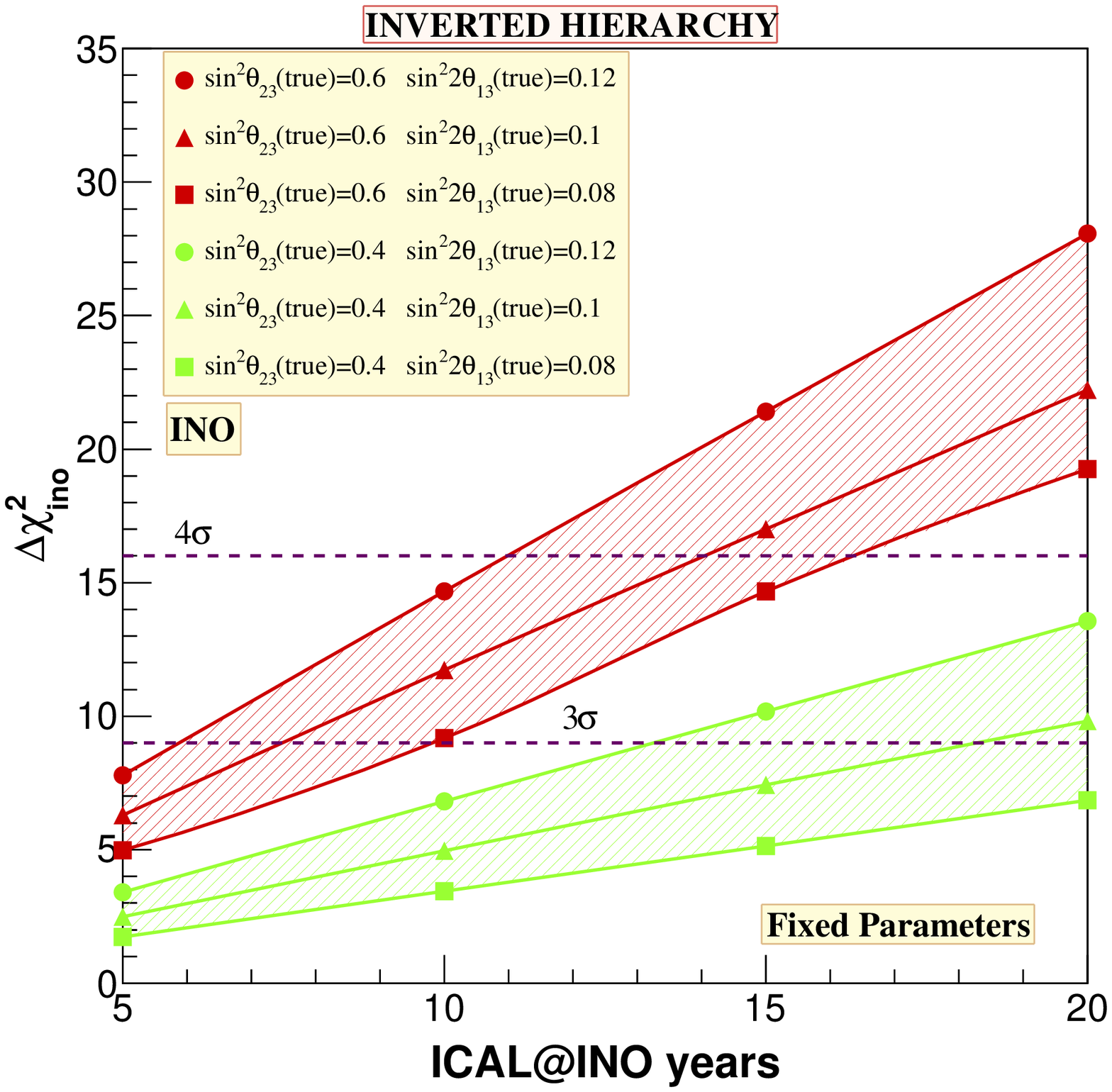}
\caption{Same as Fig. \ref{fig:hierfixed} but for $\sat=0.4$ (green band) and $\sat=0.6$ (red band).}
\label{fig:hiersafixed}
\end{figure}

\begin{figure}
\centering
\includegraphics[width=0.48\textwidth]{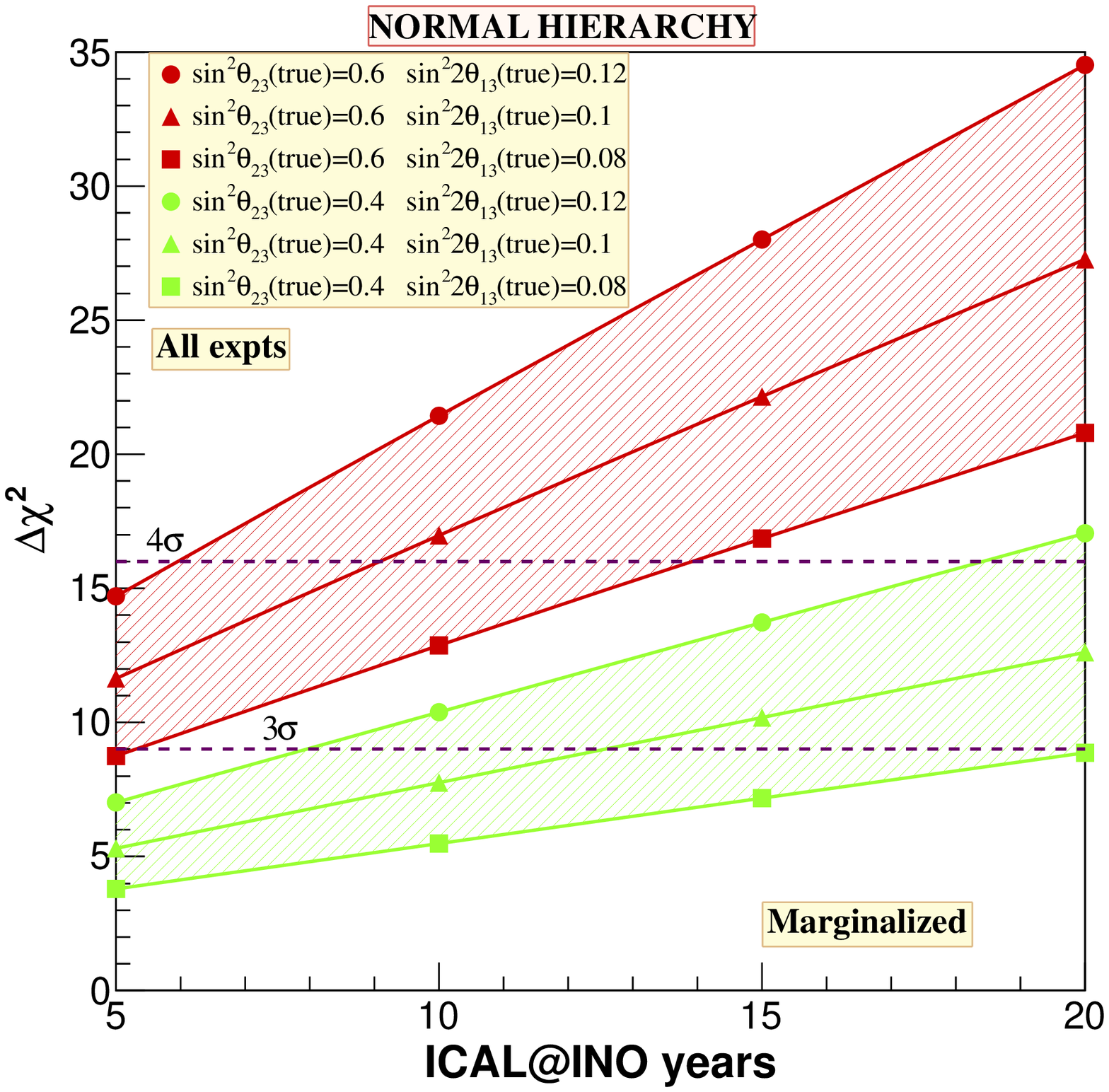}
\includegraphics[width=0.48\textwidth]{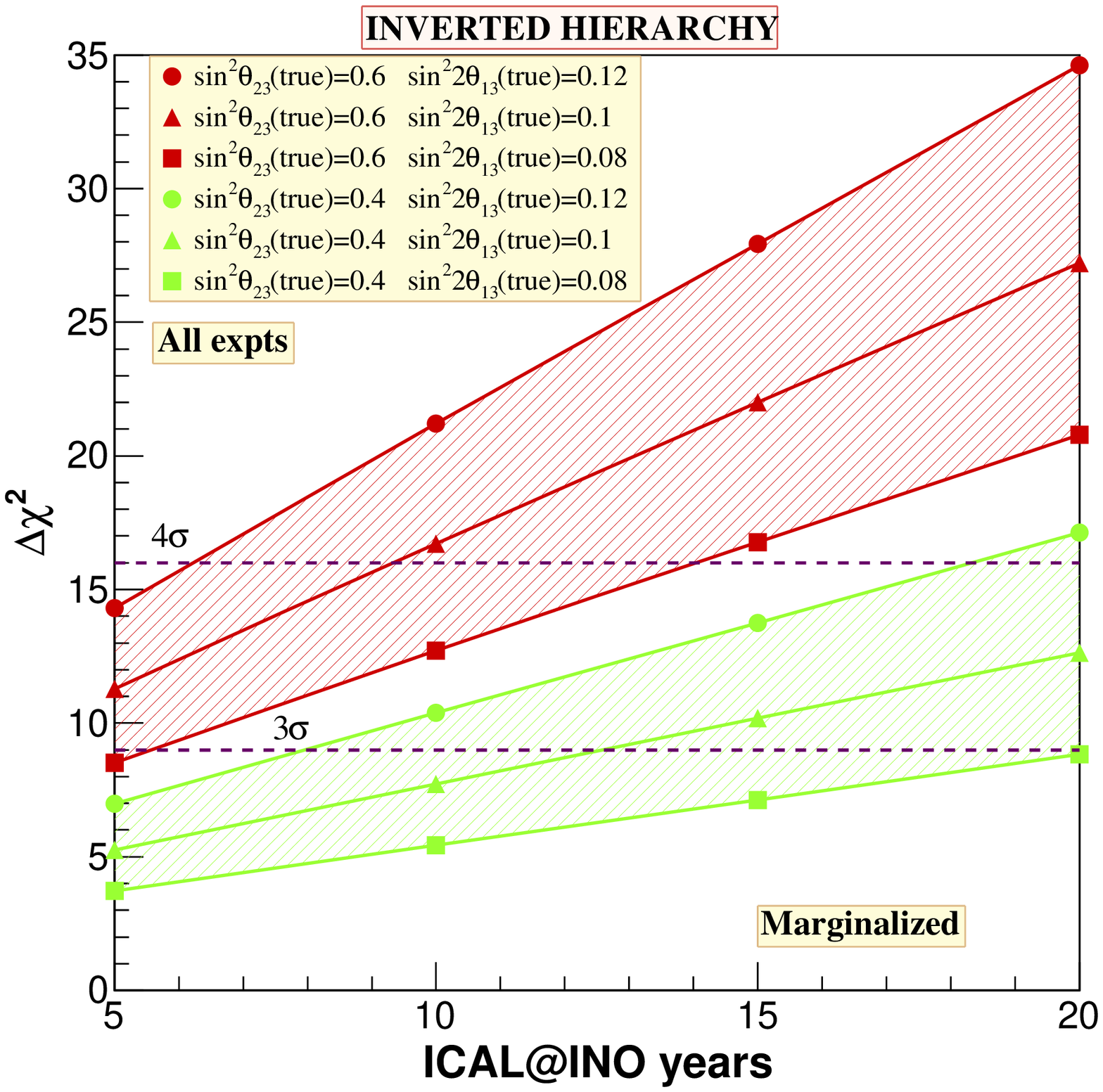}
\caption{Same as Fig. \ref{fig:hiersens} but for $\sat=0.4$ (green band) and $\sat=0.6$ (red band).}
\label{fig:hiersaall}
\end{figure}

It is well known that the amount of earth matter effects increases with increase 
in both $\theta_{13}$ and $\theta_{23}$. 
In the previous plots, we showed the mass hierarchy sensitivity 
for different allowed values of $\stcht$, while $\sat$ was fixed at 
maximal mixing. In Figs. \ref{fig:hiersafixed} and \ref{fig:hiersaall} we 
show the sensitivity to the neutrino mass hierarchy as a function of 
number of years of running of ICAL@INO for different values of 
$\stcht$ as well as $\sat$. In Fig. \ref{fig:hiersafixed} we show the 
$\Delta \chi^2$ corresponding to ICAL@INO alone and with 
oscillation parameters fixed in the fit at their true values. This figure 
corresponds to Fig. \ref{fig:hierfixed} of the previous section but now with 
two other values of $\sat$. In Fig. \ref{fig:hiersaall} we give the combined 
sensitivity to mass hierarchy of all accelerator and reactor experiments 
combined with the data of ICAL@INO. We reiterate that the 
$x$-axis of the Fig. \ref{fig:hiersaall} shows the exposure of ICAL@INO, 
while for all other experiments we have assumed the full run time as 
discussed in section \ref{sec:lbl}. 
The red bands in Figs. \ref{fig:hiersafixed} and \ref{fig:hiersaall} 
correspond to $\sat=0.6$ while 
the green bands are for $\sat=0.4$. The width of each of the bands is mapped by 
increasing the value of $\stcht$ from 0.08, through 0.1, and up to 0.12. 
As seen in the previous subsection, the $\Delta \chi^2$ for the wrong mass hierarchy 
increases with $\stcht$ for a given value of $\sat$ and 
ICAL@INO exposure. A comparison of the 
$\Delta \chi^2$ for different values of $\sat$ reveals that the $\Delta \chi^2$ also 
increases with $\sat$. 

From Fig. \ref{fig:hiersaall} one infers that for $\dcpt=0$,  
a combined analysis of all relevant experimental data including 
5 years of ICAL@INO exposure would 
give the neutrino mass hierarchy from anywhere between about $2\sigma$ to $3.8\sigma$, 
depending on the values of $\sat$ and $\stcht$. With 10 years of running of 
ICAL@INO this would improve to $2.3\sigma$ to $4.6\sigma$, depending on what 
value of $\stcht$ and $\sat$ have been chosen by mother Nature. 
Here we have 
allowed $\sat$ to vary between $[0.4-0.6]$ and $\stcht$ between $[0.08-0.12]$. 
We next look at the impact of $\dcpt$ on the prospects of determining the 
neutrino mass hierarchy.

\section{Impact of $\delta_{CP}$ and $\dcpt$}
\label{sec:delta}

\begin{figure}
\centering
\includegraphics[width=0.48\textwidth]{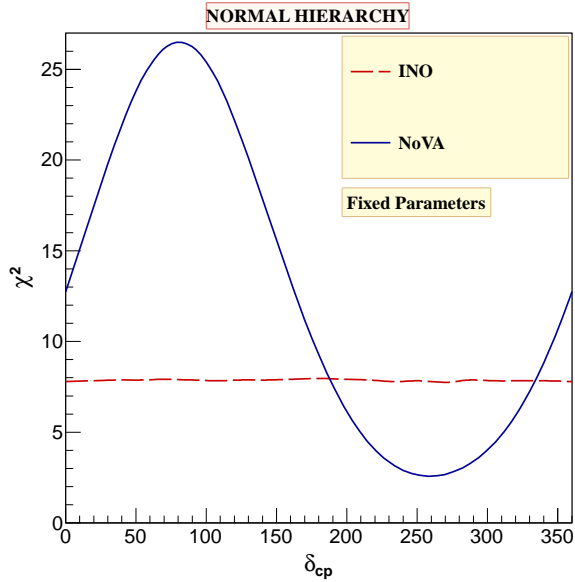}
\caption{The change of $\Delta \chi^2$ for the wrong hierarchy as a function of the 
$\delta_{CP}$ chosen in the fit. The data was generated for normal hierarchy and $\delta_{CP}$(true)=0. 
The other oscillation parameters in both data and theory are fixed at their benchmark 
values given in Table \ref{tab:param}. The solid blue line shows the change in $\Delta \chi^2$ with $\delta_{CP}$ 
for NO$\nu$A, while the dashed red line shows the corresponding 
variation of $\Delta \chi^2$ for ICAL@INO. The ICAL@INO exposure was taken as 10 years. 
}
\label{fig:hierinocp}
\end{figure}

So far we had taken $\delta_{CP}$(true)=0 in the data and 
varied $\delta_{CP}$ in the fit only for the long baseline experiments. 
For the analysis of the ICAL@INO data, we had 
kept $\delta_{CP}$ fixed to zero in both the data and the theory. 
The reason was that while the $\chi^2$ for T2K and NO$\nu$A are strongly 
dependent on $\delta_{CP}$, the mass hierarchy $\chi^2$ for 
ICAL@INO shows a very mild dependence on it. 
We show this dependence explicitly in Fig. \ref{fig:hierinocp} for 
10 years exposure in ICAL@INO and compare it with the corresponding dependence 
of NO$\nu$A (see also \cite{bs2012})\footnote{This figure was 
shown in \cite{bs2012}. However, the analysis in \cite{bs2012} 
was in terms of neutrinos done with some assumed values of the 
detector resolutions and efficiencies. Since we do here the 
complete analysis of the ICAL@INO projected data in terms of 
the detected muons and with realistic detector resolutions and 
efficiencies obtained from ICAL simulations, we reproduce a
similar plot for completeness. }. 
We generate the data for normal hierarchy and 
at the benchmark values of the oscillation parameters from 
Table \ref{tab:param} and with $\sat=0.5$ and $\stcht=0.1$. 
In the fit with inverted hierarchy, 
we keep all oscillation parameters fixed, except $\delta_{CP}$ 
which is varied over its full range $[0-2\pi]$. The corresponding $\Delta \chi^2$ is plotted in 
Fig. \ref{fig:hierinocp} as a function of the $\delta_{CP}$ in the fit. 
The red long-dashed line shows the $\delta_{CP}$ dependence of 
$\Delta \chi^2$ for ICAL@INO, while the  blue solid line shows the 
wild fluctuation of $\Delta \chi^2$ expected for NO$\nu$A. 
Amongst the accelerator and reactor experiments we show only NO$\nu$A 
in this figure as the leading contribution to the mass hierarchy comes 
from this experiment. 
We reiterate that at 
each point we use the data generated at $\delta_{CP}$(true)=0. 
The figure shows that when we fit the data with inverted hierarchy, 
the $\Delta \chi^2$ for NO$\nu$A changes from more that 26 
for $\delta_{CP}  \simeq 80^\circ$ to less than 3 for $\delta_{CP} \simeq 260^\circ$. 
When marginalized over $\delta_{CP}$ in the fit, obviously it will return the lowest 
value of the $\Delta \chi^2$, which in this case would be 2.59 \footnote{Note that here the 
other parameters are fixed and only NO$\nu$A data is being considered in the 
analysis, while the $\Delta \chi^2$s shown in Table \ref{tab:contr} are for a global fit of 
all data with all oscillation parameters allowed to vary freely in the fit and the 
combined $\chi^2$ marginalized over them.}. When marginalized over all 
oscillation paremeters, the sensitivity further reduces to $\Delta \chi^2=1.77$. 
The contribution from ICAL@INO on the other hand is seen to be 
almost independent of $\delta_{CP}$. Note that such a figure was also 
shown in \cite{choubeynu2012} for the ICAL@INO simulations. However, 
the analysis of the ICAL data has been improved since then and more 
types of systematic uncertainties introduced. This explains the change in the 
behavior of this plot with the dependence of the $\Delta \chi^2$ becoming 
flatter with $\delta_{CP}$. From this study we conclude that one does not 
need to marginalize over $\delta_{CP}$ for the ICAL@INO data. However, 
for the long baseline experiments a fine marginalization over this 
parameter is absolutely crucial, especially for mass hierarchy studies.

\begin{figure}[t]
\centering
\includegraphics[width=0.48\textwidth]{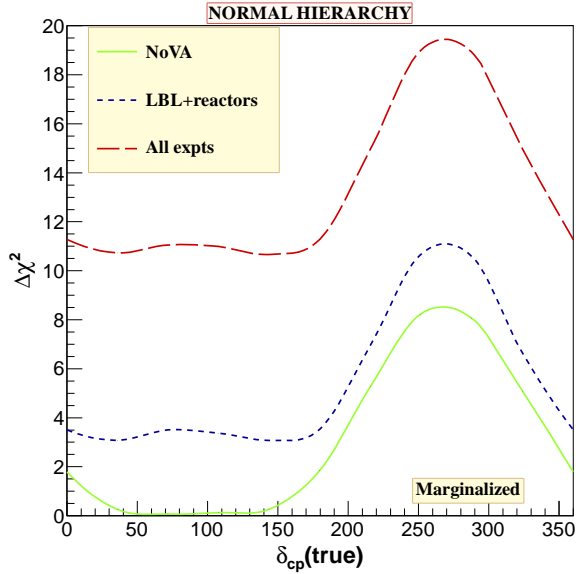}
\caption{Impact of $\delta_{CP}$(true) on the mass hierarchy sensitivity. 
The sensitivity change of NO$\nu$A due to $\dcpt$ is shown by the 
green solid line. The $\Delta \chi^2$ for the wrong mass hierarchy 
expected from the combined data from 
T2K, NO$\nu$A and the reactor experiments is shown  by the 
short-dashed blue line. The global $\Delta \chi^2$ for the wrong mass hierarchy 
from the combined ICAL@INO plus the accelerator and reactor data is shown by 
the red long-dashed line. Data was generated for normal hierarchy and at the 
benchmark oscillation point from Table \ref{tab:param} with $\sat=0.5$ and 
$\stcht=0.1$ and at each value of $\dcpt$ shown in the $x$-axis. The 
fit to the wrong inverted hierarchy was fully marginalized over all oscillation 
parameters. The ICAL@INO exposure was taken as 10 years. 
}
\label{fig:hierlblcp}
\end{figure}

We had seen in Fig. \ref{fig:hierinocp} (and as is well known) 
that the long baseline experiments are very sensitive to $\delta_{CP}$. 
In that figure we were studying the impact of changing $\delta_{CP}$ in the 
fit for a particular $\delta_{CP}$(true) in the data. 
In particular, we had taken 
$\delta_{CP}$(true)=0. 
A pertinent question at this point 
is the following: what is the impact of $\delta_{CP}$(true) on the sensitivity of the experiments 
to the neutrino mass hierarchy? We present the answer to this 
question in Fig. \ref{fig:hierlblcp}, where we show the $\Delta \chi^2$ for the mass hierarchy
sensitivity as a function of $\delta_{CP}$(true). To obtain these curves, we 
generate data for normal hierarchy at each value of $\delta_{CP}$(true) shown in the $x$-axis 
and then 
fit this data for inverted hierarchy by marginalizing over {\it all} oscillation parameters, 
including $\delta_{CP}$. The data were generated for the benchmark oscillation 
point given in Table \ref{tab:param}, $\sat=0.5$ and $\stcht=0.1$. The exposure 
considered is 10 years of ICAL@INO and full run for all other experiments. 
The green solid line in this figure is for only 
NO$\nu$A, the blue short-dashed line is obtained when we combine NO$\nu$A, T2K and 
all the reactor data, while the red long-dashed line is what we get when the ICAL@INO 
data is also added to the long baseline and reactor data. As expected, the ICAL@INO 
data is almost completely independent of $\delta_{CP}$(true) and so is its projected 
sensitivity to the neutrino mass hierarchy. 
On the other hand, the reach of the NO$\nu$A experiment for determining the 
neutrino mass hierarchy is seen to be extremely sensitive to the value of 
$\delta_{CP}$(true). All our plots shown so far on the 
global mass hierarchy reach were done assuming $\dcpt = 0$. We can see from the 
figure that indeed the statistical significance with which we could rule out the 
inverted mass hierarchy in this case is $3.4\sigma$, as discussed before. 
The $\Delta\chi^2$ for $\dcpt=0$ for NO$\nu$A is 1.77. However, this 
quickly falls to almost zero for $\dcpt \simeq [50^\circ - 150^\circ]$. Thereafter, 
it rises sharply giving a $\Delta \chi^2= 8.21$ around $\dcpt\simeq 270^\circ$, and 
then finally falls back to $\Delta \chi^2= 1.77$ for $\dcpt=360^\circ$. When 
T2K and all reactor data are added, there is an improvement to the combined sensitivity 
due to constraint coming from the mismatch between the best-fit for different experiments. 
This is specially relevant in the $\dcpt \simeq [50^\circ - 150^\circ]$ range where NO$\nu$A 
by itself gives no mass hierarchy sensitivity. However, once we add the T2K and reactor 
data to NO$\nu$A data, the $\Delta \chi^2$ of this combined fit in this region of 
$\dcpt$ increases to $\simeq 3.5$. The reason for this can be understood as follows.
For the case where $\dcpt=72^\circ$, the best-fit for NO$\nu$A alone was $\dcp=234^\circ$, 
$\sa=0.5$ and $\stch=0.1$. For this $\dcpt$, T2K data taken alone gave $\Delta \chi^2 \simeq 0$ 
with best-fit at  
$\dcp=198^\circ$, 
$\sa=0.52$ and $\stch=0.08$. 
A combined fit with all accelerator and reactor data gave best-fit at 
$\dcp=198^\circ$, 
$\sa=0.48$ and $\stch=0.1$. This results in a contribution to the mass hierarchy $\Delta\chi^2=0.92$ 
from NO$\nu$A, $\Delta\chi^2=1.41$ from T2K and $\Delta\chi^2=1.1$ from reactors. 
That the reactor data return a $\Delta \chi^2$ contribution to the mass hierarchy sensitivity 
might appear strange at the outset since the combined fit given above has a best-fit $\stch=0.1$, and 
one usually does not expect the reactor data to depend on 
sign of $\ma$, $\dcp$ and $\sa$. However, note 
that the reactor data depend on $|\ma|$ and what we use in our fits is 
$\meff$ given by Eq. (\ref{eq:meff}) which is related to $\dcp$. Therefore, 
as discussed earlier in section 6, 
this subtlety regarding the choice of the definition of the neutrino mass 
hierarchy results in a small change in the best-fit $|\ma|$, which in turn 
results in a small $\Delta \chi^2$ contribution 
to the mass hierarchy from the reactor experiments. 
Finally, addition of the ICAL@INO data raises the $\Delta \chi^2$ by a constant amount 
for all values of $\dcpt$. Therefore, depending on $\dcpt$ the combined sensitivity to 
the neutrino mass hierarchy could range from $3.3\sigma$ (for $\dcpt=144^\circ$) to 
$4.4\sigma$ (for $\dcpt\simeq 270^\circ$). These numbers are for $\sat=0.5$ and $\stcht=0.1$ and 
will improve for larger values of these parameters.

\section{Conclusions}
\label{sec:con}

In this paper we looked in detail at the prospects of determining the neutrino mass 
hierarchy with the data collected in the atmospheric neutrino experiment 
ICAL@INO. The atmospheric muon neutrino events before oscillations 
were simulated using the NUANCE 
based generator developed for ICAL@INO. To reduce Monte Carlo 
fluctuations 1000 years of exposure was used for generating the events. 
Since it takes very long for the generator to produce such a large event sample, 
we simulated the atmospheric events using the generator just once for 
no oscillations and used a  reweighting algorithm to obtain the oscillated event 
sample for any set of oscillation parameters. The oscillated muon event spectrum 
was then folded with the muon reconstruction efficiencies, charge identification 
efficiencies, energy resolution and the zenith angle resolution functions 
obtained from ICAL simulations (which will appear elsewhere \cite{inomuon}) 
to obtain the reconstructed muon event spectrum in the detector. 
As ICAL is a magnetized calorimetric detector allowing an identification of $\mu^-$ 
and $\mu^+$ events, it has
an edge over rival atmospheric neutrino experiments. 
We defined a $\chi^2$ function for Poissonian distribution for the errors in the 
ICAL@INO experiment taking into account systematic uncertainties expected 
in the experiment. 

The data was generated for benchmark true values for the oscillation 
parameters and a given neutrino mass hierarchy and fitted with the wrong hierarchy. 
We showed the mass hierarchy sensitivity results with only ICAL@INO data for the 
analysis with fixed values of the oscillation parameters 
in the fit, as well as that obtained after marginalization 
over $|\meff|$, $\sa$ and $\stch$ in their current $3\sigma$ ranges.  
We showed these results as a function of the exposure in ICAL@INO. 
From a comparison of the two results, we 
showed that the mass hierarchy sensitivity with ICAL@INO data 
deteriorates with the uncertainty in the measured value of $|\meff|$, $\sa$ and $\stch$. 
These parameters will be rather accurately determined by the T2K, NO$\nu$A 
Double Chooz, RENO and Daya Bay 
experiments. Since INO is expected to start operation 
after these have finished their full projected run, it is 
meaningful to include their effect in a 
combined statistical analysis for the neutrino mass hierarchy. 
In order to take that into account, 
we simulated the data for these experiments using GLoBES 
with the experimental specifications mentioned in their respective 
Letter Of Intent and/or Detailed Project Report. The results on mass hierarchy sensitivity 
from the combined analysis 
of data from ICAL@INO along with that from T2K, NO$\nu$A, 
Double Chooz, RENO and Daya Bay was shown for 
benchmark values of the oscillation parameters and full 
marginalization over all oscillation parameters in the fit for the 
wrong mass hierarchy. We showed that 
marginalization over $\dcp$ is practically unessential for the 
ICAL@INO data. However, for the accelerator data it is absolutely crucial 
to marginalize over $\dcp$ due to the very strong dependence of the 
hierarchy sensitivity on this parameter in these experiments. 
We then generated the data at all values of $\dcpt$ and 
showed that the mass hierarchy sensitivity of ICAL@INO was independent of $\dcpt$, 
however, the sensitivity of the combined NO$\nu$A, T2K and the reactor experiments 
depends very strongly on what $\dcpt$ has been chosen by Nature. 
For $\sat=0.5$ and $\stcht=0.1$ the combined data of 
10 years exposure in ICAL@INO along with T2K, NO$\nu$A and 
reactor experiments could rule out the wrong hierarchy with a statistical 
significance of $3\sigma$ to $4.2\sigma$, depending on the chosen value 
of $\dcpt$. 
We also studied the effect of $\stch$(true) and 
$\sa$(true) on the reach of these 
combined projected data sets to determining the neutrino mass 
hierarchy. For $\dcpt=0$, we showed that the statistical significance with which 
the wrong hierarchy could be ruled out by the global data set comprising of 
10 years exposure in ICAL@INO along with 
T2K, NO$\nu$A and reactor experiments, could be anywhere between 
$2.13\sigma$ to $4.5\sigma$ depending on $\sat$ and $\stch$, where we 
allowed $\sat$ to vary between $[0.4-0.6]$ and $\stcht$ between $[0.08-0.12]$. 
For the most favorable choice of $\dcpt\simeq 270^\circ$ the sensitivity 
could go up to greater than $5\sigma$ with 10 years of 
ICAL@INO combined with 
data from T2K, NO$\nu$A and reactor experiments. 

\section{Appendix -- Impact of Definition of Mass Hierarchy
\label{sec:appendix}}

The first 
row of Table \ref{tab:contr} 
shows the $\Delta \chi^2$ for the case used in this paper where 
normal hierarchy is defined as $\meff >0$ and inverted hierarchy 
is defined as $\meff <0$. In this case the survival probability $P_{\numu\numu}$ 
is almost the same for the normal and inverted hierarchies when $\theta_{13}=0$. 
However, this is not true for the channels $P_{\numu \nue}$ relevant for 
T2K and NOvA and $P_{\nue\nue}$ relevant for the reactor experiments. 
Thus in the best-fit, the value of $|\ma|$ and $\dcp$ has to be suitably adjusted 
so that the oscillation probabilities in these oscillation channels 
are closest to each other in the data and in the fit. This results in 
a small $\Delta \chi^2$ from experiments that by themselves have practically no 
hierarchy sensitivity at all. We have indeed checked that the 
$\Delta \chi^2$ for mass hierarchy is zero for all reactor experiments and T2K, when 
the fit is performed using data from only one experiment at a time and all 
oscillation parameters allowed to vary in the fit. 
Only when one performs a combined analysis does this tension between choice of 
the oscillation frequencies arise, returning a small contribution to 
the mass hierarchy from the reactor experiments and T2K. 

In order to check the impact of our choice of $\meff >0$ ($\meff <0$) as 
normal (inverted) hierarchy, we repeated our global fit by taking $\ma >0$ 
as the definition for normal hierarchy and $\ma <0$ as the definition for 
inverted hierarchy. The result for this case is shown in the second row 
of Table \ref{tab:contr}. One can see that there is hardly any impact of 
the definition of mass hierarchy on our final results.

\vskip 1cm
\noindent
{\Large \bf Acknowledgements}
\vskip 0.4cm

\noindent
This work is a part of the ongoing effort of INO-ICAL collaboration
to study various physics potential
of the proposed INO-ICAL detector. Many members of the collaboration
have contributed for the completion
of this work. We would like to specially mention the contribution of
Gobinda Majumder and Asmita Redij
for developing the INO-ICAL GEANT-4 detector simulation and
reconstruction packages,  A. Chatterjee, K. K. Meghna and K. Rawat for producing
the muon resolutions and efficiencies in the ICAL detector, and
A. Dighe, A. Chatterjee, P. Ghoshal, S. Goswami, D. Indumathi, N. Mondal,  
Md. Nayeem, S. Uma Sankar and N. Sinha  for participating in regular discussions
in the course of this work. We are very grateful to A. Dighe, V. Datar, A. Raychaudhuri, 
N. Mondal and M.~V.~N.~Murthy  
for critical reading of the manuscript. S.C. acknowledges partial support from the European 
Union FP7 ITN INVISIBLES (Marie Curie Actions, PITN-GA-2011-289442).


\end{document}